\documentclass[preprint,12pt,authoryear]{elsarticle}

\usepackage{amssymb}

\usepackage{subcaption}
\usepackage{wrapfig,caption}
\usepackage{multirow}
\usepackage{amsmath}
\usepackage{upgreek}
\usepackage{hyperref}
\usepackage{fancyhdr}
\usepackage{enumitem}
\usepackage{color}
\usepackage{array}
\usepackage{amstext}
\usepackage{amsmath} 
\usepackage{amssymb}
\usepackage[usenames,dvipsnames,svgnames,table]{xcolor}
\usepackage{psfrag}
\usepackage{array}
\usepackage{threeparttable}
\usepackage{dcolumn}
\newcolumntype{d}{D{.}{.}{-1}}
\usepackage{enumitem}
\usepackage{titlesec}
\usepackage{float}
\usepackage{comment}

\newcommand\Reyn{\mbox{\text{Re}}}  
\newcommand\Pran{\mbox{\text{Pr}}}  
\newcommand\Pec{\mbox{\text{Pe}}}  
\newcommand\deriv{\mbox{\text{d}}}  

\newcommand\Peclet{\mbox{{P\'{e}clet}}}

\usepackage{xcolor}
\usepackage{tikz}
\usepackage{pgfplots}
\usepackage{bm}

\pgfplotsset{
        layers/my layer set/.define layer set={
            background,
            main,
            foreground
        }{
        },
        set layers=my layer set,
    }

\definecolor{findOptimalPartition}{HTML}{D7191C}
\definecolor{storeClusterComponent}{HTML}{FDAE61}
\definecolor{dbscan}{HTML}{ABDDA4}
\definecolor{constructCluster}{HTML}{2B83BA}

\definecolor{ColorOne}{HTML}{61BB46}
\definecolor{ColorTwo}{HTML}{FDB827}
\definecolor{ColorThree}{HTML}{F5821F}
\definecolor{ColorFour}{HTML}{E03A3E}
\definecolor{ColorFive}{HTML}{963D97}
\definecolor{ColorSix}{HTML}{009DDC}
\definecolor{Invisible}{HTML}{FFFFFF}

\usepackage{tikz}
\usepackage{blindtext}
\usetikzlibrary{plotmarks}

\usepackage{pgfplots}
\pgfplotsset{compat=newest}

\usetikzlibrary{plotmarks}
\usepackage{array,booktabs}

\usepackage{scalerel}
\usepackage{stackengine,wasysym}

\newcommand\reallywidetilde[1]{\ThisStyle{%
  \setbox0=\hbox{$\SavedStyle#1$}%
  \stackengine{-.1\LMpt}{$\SavedStyle#1$}{%
    \stretchto{\scaleto{\SavedStyle\mkern.2mu\AC}{.5150\wd0}}{.6\ht0}%
  }{O}{c}{F}{T}{S}%
}}

\newcommand{\uncorrf}{\raisebox{1pt}{\tikz{\draw[thick, violet, solid](0,0)--(4mm,0);}}}
\newcommand{\uncorrp}{\raisebox{1pt}{\tikz{\draw[thick, blue, solid](0,0)--(4mm,0);}}}

\newcommand{\corrf}{\raisebox{1pt}{\tikz{\draw[thick, violet, densely dotted](0,0)--(4mm,0);}}}
\newcommand{\corrp}{\raisebox{1pt}{\tikz{\draw[thick, blue, densely dotted](0,0)--(4mm,0);}}}

\newcommand{\ELf}{\raisebox{1pt}{\tikz{\draw[thick, violet, dashed](0,0)--(4mm,0);}}}
\newcommand{\ELp}{\raisebox{1pt}{\tikz{\draw[thick, blue, dashed](0,0)--(4mm,0);}}}

\journal{International Journal of Heat and Mass Transfer}

\begin{document}

\begin{frontmatter}



\title{On the thermal entrance length of moderately dense gas-particle flows}


\author[label1]{S. Beetham}
\author[label1]{A. Lattanzi}
\author[label1,label2]{J.Capecelatro}
\address[label1]{University of Michigan, Department of Mechanical Engineering}
\address[label2]{University of Michigan, Department of Aerospace Engineering}

\begin{abstract}
The dissipative nature of heat transfer relaxes thermal flows to an equilibrium state that is devoid of temperature gradients. The distance to reach an equilibrium temperature -- the thermal entrance length -- is a consequence of diffusion and mixing by convection. The presence of particles can modify the thermal entrance length due to interphase heat transfer and turbulence modulation by momentum coupling. In this work, Eulerian--Lagrangian simulations are utilized to probe the effect of solids heterogeneity (e.g., clustering) on the thermal entrance length. For the moderately dense systems considered here, clustering leads to a factor of 2--3 increase in the thermal entrance length, as compared to an uncorrelated (perfectly mixed) distribution of particles. The observed increase is found to be primarily due to the covariance between volume fraction and temperature fluctuations, referred to as the fluid drift temperature. Using scaling arguments and Gene Expression Programming, closure is obtained for this term in a one-dimensional averaged two-fluid equation and is shown to be accurate under a wide range of flow conditions.

\end{abstract}

\end{frontmatter}

\section{Introduction}
\label{sec:intro}
Internal flow exhibiting purely dissipative heat transfer exchanges heat with walls or its surroundings and its temperature profile relaxes to an equilibrium. The thermal entrance length, $l_{\text{th}}$, is defined as the length after which temperature gradients with respect to non-homogeneous directions vanish. Over the last several decades, the thermal entrance length has been studied extensively in the context of laminar and turbulent single-phase flows \cite[see, e.g.,][]{Sparrow1957, Abbrecht1960, Lee1968, Hasan2012}. For laminar flow bounded by constant temperature walls, the entrance length may be estimated by~\citep{incropera_2011}
\begin{equation}\label{eq:lth-single}
   l_{\text{th}}/D = 0.05 \Reyn_{{D}} \Pran, 
\end{equation}
where $\Reyn_{{D}}$ is the Reynolds number characterized by the vessel (pipe/duct) width or diameter, $D$, and $\Pran$ is the Prandtl number. When considering heat transfer in turbulent flows, the Nusselt number is used to assess when the flow has reached a fully-developed state. In particular, the entrance length is defined as the length after which the Nusselt number is within several percent (typically 1 to 5\%) of the fully developed value and can be thought of as the thermal equivalent to a hydrodynamic boundary layer. While values across this range are used throughout the literature, \citet{Sparrow1957} pointed out that 5\% has more utility for comparison with experimental results, where achieving accuracy within 1 or 2\% is challenging. Several models have been proposed for the Nusselt number in recent years, based upon the thermally evolving, turbulent pipe flow. Many draw upon the Dittus--Boelter correlation~\citep{Welty2019}, given as Nu$= 0.023\text{Re}_D^{0.8}\text{Pr}^{n}$ where $n=0.4$ for a heated fluid and 0.3 if it is cooled. Several other works also formulate dependencies upon the nondimensional length scale $L/D$ \cite[e.g.,][]{Hasan2012}, where $L$ is the pipe length.

While thermally evolving and wall-bounded, single-phase flows are of great importance (e.g., cooling systems for nuclear reactors, tube heat exchangers, etc.), many applications of interest also contain a disperse phase that exchanges heat with the fluid. Of particular interest in this work are turbulent and thermally evolving \emph{gas--solid} flows. This class of flows is pervasive in nature and industry, spanning applications from volcanic eruptions \citep{Xu2002, Wilson1978,Lube2020} to the storage of thermal energy \citep{Pouransari2017,Morris2016, Mehos2017, Pielichowska2014} and the upgrading of feedstock to usable fuels in circulating fluidized bed (CFB) reactors. In the case of CFB reactors, cool feedstock particles are fluidized with a hot gas, with the goal of mixing the phases in such a way that the hot, fluidizing gas exchanges heat with the particles, thereby initiating their devolatilization into usable fuels. In both experimental and computational studies, it has been observed that particles spontaneously organize into coherent structures (clusters), thereby reducing contact between the phases, impeding mixing and delaying heat transfer. 

Early experimental work in the 1990s by \citet{Louge1993, Ebert1993} showed that heat transfer between the particles in a fluidized bed and surrounding walls is increased by up to an order of magnitude when compared to single-phase turbulent flow. This increase in heat transfer was even more marked in denser regions of particles. In the context of a CFB reactor, \cite{Noymer1998} found that dilute suspensions of particles have the opposite effect and can impede heat transfer to the wall. In addition to these experimental works, several contemporary computational works have demonstrated that coherent structuring of particles may inhibit mixing between phases and detrimentally impact heat transfer \citep{agrawal2013filtered,miller2014carbon,pouransari2017effects,Guo2019,Beetham2019}. This phenomenon has important implications for reactor design and efficiency, since reduced heat transfer performance impacts the thermochemical conversion rate. Despite the thermal entrance length's crucial role in properly sizing industrial unit operations, the effect of solids heterogeneity on this quantity remains largely unknown. 

In the last decade, advancements in high performance computing has allowed for increased access to high-fidelity and large-scale computational studies of complex multiphase flows. 
For example, \citet{Lei2020} use progressive filtering of highly resolved simulations to formulate an improved model for the interphase heat transfer coefficient. \citet{Rauchenzauner2020} derive a spatially averaged Euler--Euler model for heat transfer for wall-bounded, dense gas--solid flows by proposing a drift temperature that represents the fluid temperature fluctuations seen by the particles, the primary quantity of interest in the present work, and propose closures.  \citet{Jofre2020} studied heat transfer in irradiated turbulent dilute gas--solid flow using a two-step approach similar to the one undertaken in this work. They determined that the residence time and structure of particles play dominant roles in describing heat transfer. Another recent work \citep{Peng2019} demonstrated that the pseudo-turbulent heat flux that arises in filtering the heat equation, is an important factor describing thermal properties. \citet{Yousefi2021} employed particle-resolved direct numerical simulation to probe the heat transfer in particle-laden channels at moderate volume fractions, demonstrating that the turbulent heat flux dominates large scale thermal behavior.

While research on multiphase heat transfer is active and growing, the effect of particle heterogeneity on the thermal entrance length remains an open question. In this work, the thermal entrance length is examined via Eulerian--Lagrangian simulations by employing a two-step approach. First a moderately dense isothermal, gas-particle flow is simulated to generate realistic clustering. Next, the cold-flow simulations are fed into a statistically one-dimensional domain with a prescribed temperature difference between the phases at the inlet boundary. From these results, we quantify the effect of mean solids volume fraction, \Peclet\; number and ratio of heat capacities on the thermal entrance length and propose scaling relations for both clustered and uniform gas--solid flows. We then derive the two-fluid heat equations, quantify which terms are responsible for deviations from an uncorrelated solids phase and propose a model for the dominant, unclosed term. This model is formulated using both scaling arguments and Gene Expression Programming and is compatible with existing two-fluid theory. 

\section{System configuration} 
\label{sec:problemStatement}

\begin{figure}[ht]
    \centering
    \includegraphics[width=\textwidth]{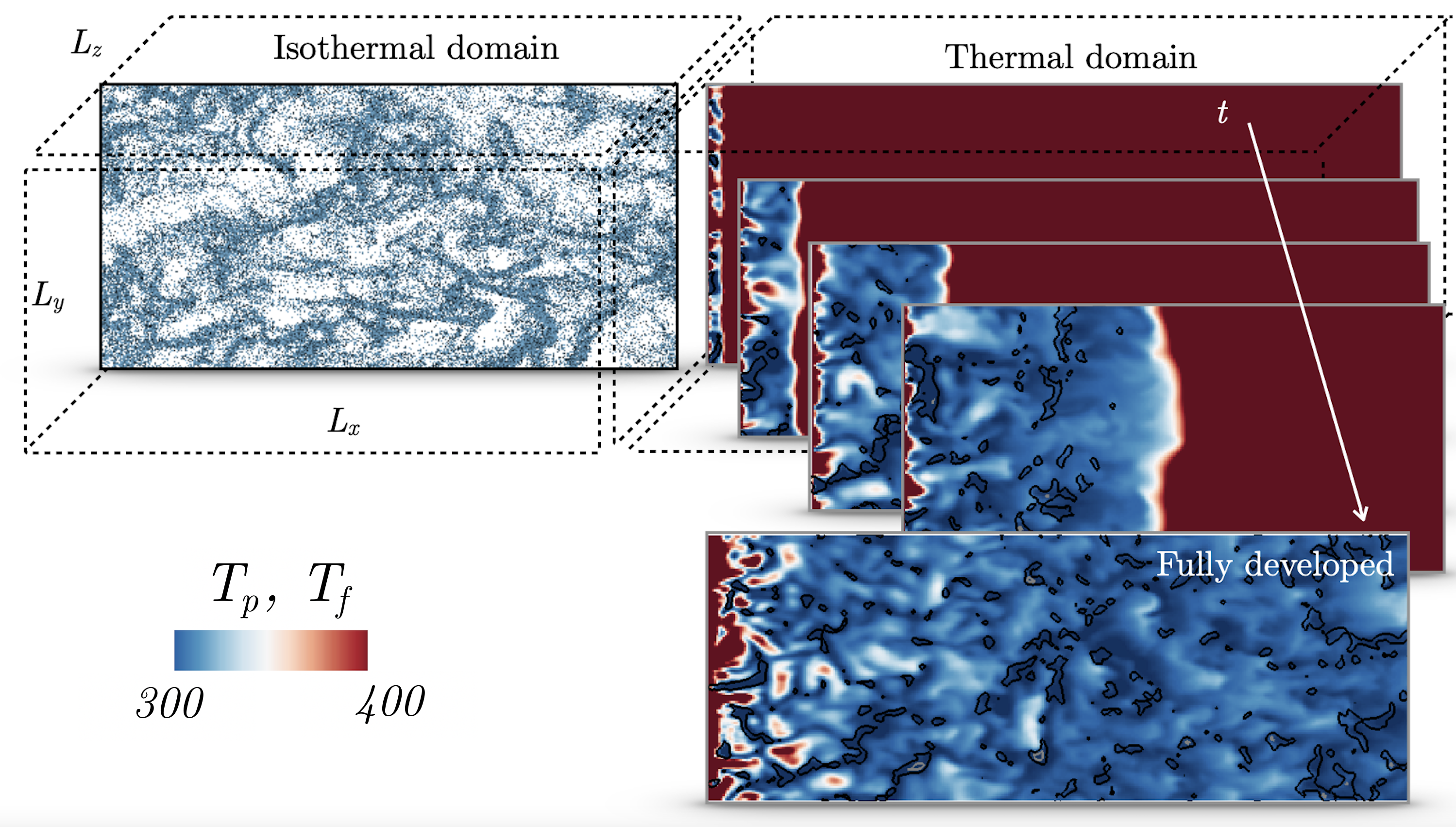}
    \caption{A fully developed configuration of particles (cold) and gas (hot) is injected into the thermal domain. Here, the initially cold particles are shown in the left pane. On the right, instantaneous snapshots of the fluid phase temperature is shown from an early time (top) to a fully developed period (bottom).}
    \label{fig:setup}
\end{figure}

In this work, our goal is to examine the effect of realistic multiphase hydrodynamics on heat transfer and thermal entrance length. To do this, we use a two-step simulation setup representative of the fully-developed interior of a riser in a CFB reactor. A sketch of this configuration is outlined in Fig.~\ref{fig:setup}. Here, clustering behavior is established in an isothermal simulation, which then is fed into a thermally-evolving domain.  

\subsection{Isothermal simulations} 
\label{subsec:isothermal}
Prior to simulating thermally-evolving two-phase flows, the hydrodynamics are established in a separate set of simulations. We consider three-dimensional, homogeneous, fully-developed gas--solid riser flow in the absence of heat transfer. In these simulations, $N_p$ particles each with diameter $d_p$ and density $\rho_p$ are initially randomly distributed in a quiescent gas with density $\rho_f$ and kinematic viscosity $\nu_f$.  A body force (gravity, $g$) drives the hydrodynamics and the mass flow rate is forced such that the mean fluid velocity is held at a fixed value, $u_{\text{bulk}}$, mimicking the flow conditions inside a riser. Here, $u_{\text{bulk}}$ exceeds the anticipated particle settling velocity, and opposes the direction of gravity such that the particles are entrained in the fluid. 

Due to strong coupling between the phases, the particles form dense clusters that generate correlation between the particle volume fraction, $\varepsilon_p$, and fluid velocity, $u_f$. When in a correlated (clustered) configuration, assemblies of particles experience enhanced settling, on the order of 2 to 3 times greater than the terminal velocity of an isolated particle, $\mathcal{V}_0$. This increased settling establishes a mean slip velocity between the phases that is not known \emph{a priori} \cite[see][for more details]{capecelatro2014cit, Beetham2021}. 

In this configuration, relatively few non-dimensional groups arise. These include the Galileo number, ${\rm Ga} = \sqrt{(\rho_p/\rho_f -1)d_p^3 g}/\nu_f$; the bulk Reynolds number, ${\rm Re}_{\text{bulk}} = u_{\text{bulk}} d_p/\nu_f$; and the mean mass loading $\varphi=\rho_p\langle\varepsilon_p\rangle/(\rho_f\langle\varepsilon_f\rangle)$. Here, $\varepsilon_f=1-\varepsilon_p$ is the fluid-phase volume fraction, and angled brackets denote an average in all three spatial directions and time. The parameters associated with the isothermal simulations are summarized in Table~\ref{tab:parameters1}, where sets of values denote quantities that are varied in the simulations. Further details on the set up and flow physics of these simulations can be found in~\citet{Beetham2021}. 

\begin{table}[ht]
\centering
\begin{tabular}{l l r r} 
\midrule
\multicolumn{4}{c}{\textit{Dimensional quantities}} \\
\midrule
$d_p$ & Particle diameter & 90 & $\lbrack \mu \text{m} \rbrack$ \\
$\rho_p$ & Particle density & 1000 & [kg/m$^3$] \\
$\rho_f$ & Fluid density & 1 & [kg/m$^3$] \\
$\nu_f$ & Fluid viscosity & 1.8$\times$10$^{-5}$ & [kg/m s] \\ 
$u_{\text{bulk}}$ & {Bulk fluid velocity} & (0.42, 2.11, 2.95) & [m/s] \\
$g$ & Gravity & 0.8 & [m/s$^2$] \\
$\tau_p$ & Stokes response time & 0.025 & [s] \\
\midrule
\multicolumn{4}{c}{\textit{Non-dimensional Quantities}} \\
\midrule
$N_p$ & Number of particles &  \multicolumn{2}{r}{(610,370, \;15,564,442, \;30,518,514)} \\
$\varphi$ & Mass loading &  \multicolumn{2}{r}{(1.0, 26.2, 52.6)} \\ 
Ga & Galileo number &  \multicolumn{2}{r}{2.3} \\
$\langle \varepsilon_p \rangle$ & Mean particle volume fraction & \multicolumn{2}{r}{(0.001, 0.0255, 0.05)} \\
$\Reyn_{\text{bulk}}$ & Reynolds number &  \multicolumn{2}{r}{(2.1, 10.5, 14.7)} \\
\midrule
\multicolumn{4}{c}{\textit{Computational quantities}} \\
\midrule
 & Domain size ($W \times L \times L $) & $0.158 \times 0.038 \times 0.038$ & [m]\\
& Grid size ($n_x \times n_y \times n_z$) & $512\times128\times128$ & \\
\midrule
\end{tabular} 
\caption{Summary of relevant parameters for the isothermal simulations.}
\label{tab:parameters1}
\end{table}

\subsection{Thermal simulations}
\label{sec:thermal}

Once a statistically stationary state is reached in time, a snapshot of the isothermal simulation is then fed into the thermal domain. This domain is initially comprised of fluid with heat capacity, $C_{p,f}$, and thermal diffusivity, $\kappa_f$. At the inlet, the fluid temperature is given a uniform value, $T_{f,0}$. In the spanwise directions, periodic boundary conditions are employed and the domain lengths match the isothermal simulation.  In the streamwise direction, $y-z$ plane data is taken incrementally from the isothermal snapshot and fed in as an inlet condition at $x=0$. Particles are assigned a uniform temperature, $T_{p,0}<T_{f,0}$, and heat capacity, $C_{p,p}$. After injection, two-way coupling drives the phases toward thermal equilibrium. 

The thermal simulations introduce three additional relevant dimensionless groups: the Prandtl number, $\Pran = C_{p,f} \nu_f \rho_f /\kappa_f$; the P\'{e}clet number, $\Pec = d_p u_{\text{bulk}}\rho_f C_{p,f}/\kappa_f$; and the ratio of heat capacities, $\chi = C_{p,p}/C_{f,p}$. The parameters used in these simulations are summarized in Table~\ref{tab:parameters2}, where sets of values are provided for the quantities that are varied in the simulations. Using the riser of a CFB reactor as our motivation, we prescribe the inflow velocity, $u_{\text{bulk}}$, such that the resultant P\'{e}clet number corresponds to typical riser conditions \citep{Shah2016}. Three values of $\chi$ are considered, corresponding to the heat capacities of sand \citep{Hamdhan2010}, Zeolite 4A \citep{Qiu2000} (a catalyst used in the processing of conventional oil) and bagasse \citep{lathouwers2001modeling} (a woody pulp biproduct of the commercial processing of sugarcane commonly used in biomass pyrolysis). 

\begin{table}[ht]
\centering
\begin{tabular}{l l r r} 
\midrule
\midrule
\multicolumn{4}{c}{\textit{Particle-phase quantities}} \\
\midrule
 {$C_{p,p}$} & {Particle heat capacity} & (840, 921 2300) &[J/kg K] \\
$T_{p,0}$ & {Initial particle temperature} & 300 & [K] \\
\midrule
\multicolumn{4}{c}{\textit{Fluid-phase Quantities}} \\
\midrule
$C_{p,f}$ &{Fluid heat capacity} & 1.013 & [kJ/kg K] \\
$T_{f,0}$ & {Initial fluid temperature} & 400 & [K]\\
$\kappa_{f}$ & {Fluid thermal conductivity}& 0.0334 & [J/m s K]\\
\midrule
\multicolumn{4}{c}{\textit{Non-dimensional quantities}} \\
\midrule
$\Pran$  &Prandtl number & \multicolumn{2}{r}{0.7}  \\ 
$\Pec$ & P\'{e}clet number &\multicolumn{2}{r}{(1, 5, 7)} \\
$\chi$ & Ratio of heat capacities & \multicolumn{2}{r}{(829, 909, 2270)} \\
\midrule
\midrule
\end{tabular} 
\caption{Summary of parameters for the thermally evolving simulations.}
\label{tab:parameters2}
\end{table}

\section{Computational methodology} 
\label{sec:mathematicalDescription}
The numerical simulations are solved in a volume-filtered Eulerian--Lagrangian framework  for an incompressible viscous fluid with spherical, rigid particles undergoing heat exchange~\citep{Capecelatro2013,Guo2019}. The volume-filtered continuity equation is given by
\begin{equation} 
\frac{\partial}{\partial t} \left( \varepsilon_f \rho_f \right) + \nabla \cdot \left( \varepsilon_f \rho_f \bm{u}_f \right) = 0, 
\end{equation}
where $\bm{u}_f=[u_f,v_f,w_f]^T$ is the fluid velocity. In this work, the fluid-phase density $\rho_f$ is held constant. This approximation simplifies the modeling exercise in later sections and is justified based on the moderate temperature gradient imposed between the phases (between 300K and 400K). 
The fluid-phase momentum equation is given by
\begin{equation}
\frac{\partial}{\partial t} \left( \varepsilon_f \rho_f \bm{u}_f \right) + \nabla \cdot \left( \varepsilon \rho_f \bm{u}_f \otimes \bm{u}_f \right) = \nabla \cdot \bm{\tau}_f + \varepsilon_f \rho_f \bm{g} + \bm{\mathcal{F}}_{\text{inter}} + \bm{F}_{\text{mfr}}.
\end{equation} 
Here, $\bm{g}$ is gravity, $\mathcal{F}_{\text{inter}}$ is the interphase momentum exchange (which will be defined later) and $\bm{F}_{\text{mfr}}$ is an additional source term for the cold simulations to enforce a mass flow rate in the gravity-aligned direction such that the flow reaches a statistically steady state \cite[see][for details]{capecelatro2014cit}. The filtered stress tensor, $\bm{\tau}_f$ is
\begin{equation}
 \bm{\tau}_f = -p_f \mathbb{I} + \rho_f\nu_f \left \lbrack \nabla \bm{u}_f + \nabla \bm{u}^{\text{T}} - \frac{2}{3} \left( \nabla \cdot \bm{u}_f \right) \mathbb{I} \right \rbrack,
\end{equation} 
where $\mathbb{I}$ is the identity tensor and $p_f$ is pressure. Finally, conservation of energy is given as 
\begin{equation}
\rho_f C_{p,f} \frac{\partial}{\partial t} \left( \varepsilon_f T_f \right) + \rho_f C_{p,f} \nabla \cdot \left( \varepsilon_f  \bm{u}_f T_f \right) =  \kappa_f \nabla^2 T_f + \mathcal{Q}_{\text{inter}}, \label{eq:fluidheat}
\end{equation}  
where $T_f$ is the fluid temperature and $\mathcal{Q}_{\text{inter}}$ is the interphase heat exchange (defined later).  In this work, the thermal conductivity, $\kappa_f$ is specified by maintaining a constant Prandtl number, ${\rm Pr}= 0.7$.  

Particles are tracked individually in a Lagrangian manner according to Newton's second law of motion, given by 
\begin{equation}
\frac{\text{d} \bm{x}_p^{(i)}}{\text{d} t} = \bm{u}_p^{(i)}  \label{eq:newton}
\end{equation}
and 
\begin{equation} 
m_p \frac{\text{d} \bm{u}_p^{(i)}}{\text{d} t} =  \bm{f}_{\text{inter}}^{(i)} + \mathbb{C}^{(i)} + m_p \bm{g}, \label{eq:partmom}
\end{equation}
where $\bm{x}_p^{(i)}$ and $\bm{u}_p^{(i)}$ are the position and velocity of particle $i$, respectively, $m_p$ is the particle mass and $V_p$ is its volume. Square brackets denote a fluid quantity interpolated to the center position of particle $i$. The interphase transfer term is defined as 
\begin{equation}
\bm{f}_{\text{inter}}^{(i)} = V_p \nabla \cdot \bm{\tau}_f\lbrack \bm{x}^{(i)}_p \rbrack + m_p \frac{\varepsilon_f}{\tau_p} \left( \bm{u}_f\lbrack \bm{x}_p^{(i)} \rbrack - \bm{u}_p^{(i)} \right) F(\varepsilon_f, \Reyn_p)
\end{equation}
where $\tau_p = \rho_p d_p^2/(18\rho_f\nu_f)$ is the Stokes response time, $\Reyn_p$ is the particle Reynolds number, given by 
\begin{equation}
\Reyn_p = \frac{\varepsilon_f \vert \bm{u}_f \lbrack \bm{x}_p^{(i)} \rbrack - \bm{u}_p^{(i)} \vert d_p}{\nu_f},
\end{equation} 
and $F(\varepsilon_f, \Reyn_p)$ is a non-dimensional correction factor to account for volume fraction and Reynolds number effects \citep{Tenneti2011}. The force due to inter-particle collisions, $\mathbb{C}$, is modeled using a modified soft-sphere approach originally proposed by \citet{Cundall1979}. Particles are treated as inelastic and frictional with coefficients of restitution of $0.9$ and friction coefficient of $0.1$. For additional information, see \citet{Capecelatro2013}. 
The solid-phase energy equation is given by 
\begin{equation}
m_p C_{p,p} \frac{\text{d} T_p^{(i)}}{\text{d}t} =  q_{\text{inter}}^{(i)}, \label{eq:partheat}
\end{equation}
where $T_p^{(i)}$ is the temperature of the $i$-th particle, and $q_{\text{inter}}^{(i)}$ is the interphase particle heat exchange, given as 
\begin{equation}
q_{\text{inter}}^{(i)} = V_p\kappa_f \nabla^2 T_f \lbrack \bm{x}_p^{(i)} \rbrack + \frac{6 V_p \kappa_f \text{Nu}}{d_p^2} \left( T_f \lbrack \bm{x}_p^{(i)} \rbrack - T_p^{(i)} \right).
\end{equation}
Here, ${\rm Nu}$ is the $\varepsilon_p$- ${\rm Re}_p$-dependent Nusselt number correlation recently developed by \citet{Sun2016}.

The fluid-phase equations contain several interphase exchange terms that require Lagrangian information be projected to the Eulerian grid. This is accomplished by employing the two-step filtering approach described in \citet{Capecelatro2013}, in which particle data is first extrapolated to the nearest grid points, followed by a `smoothing' operation that is performed implicitly, such that the final support of the filtering operation is tied to a chosen filter size, $\delta_f$. Here, we consider a Gaussian filter kernel, $\mathcal{G}$, with $\delta_f = 7 d_p$. 
With this, interphase exchange terms are given by
\begin{equation}
\varepsilon_f = 1- \sum_{i=1}^{N_p} \mathcal{G}\left( \vert \bm{x} - \bm{x}_p^{(i)} \vert\right) V_p,
\end{equation}
\begin{equation}
\bm{\mathcal{F}}_{\text{inter}} = - \sum_{i=1}^{N_p} \mathcal{G}\left( \vert \bm{x} - \bm{x}_p^{(i)} \vert\right) \bm{f}_{\text{inter}}^{(i)} , \label{eq:Finter}
\end{equation}
and
\begin{equation}
\mathcal{Q}_{\text{inter}} = - \sum_{i=1}^{N_p} \mathcal{G}\left( \vert \bm{x} - \bm{x}_p^{(i)} \vert\right) q_{\text{inter}}^{(i)}.  \label{eq:Qinter}
\end{equation}

The equations are solved in NGA~\citep{desjardins2008high}, a fully conservative, low-Mach number finite volume solver. A pressure Poisson equation is solved to enforce continuity via fast Fourier transforms in all three periodic directions (in the isothermal simulations) and a multigrid solver is used for the thermal simulations, which are only periodic in the spanwise directions. The fluid equations are solved on a staggered grid with second-order spatial accuracy and advanced in time with second-order accuracy using the semi-implicit Crank--Nicolson.  Lagrangian particles are integrated using a second-order Runge--Kutta method. Fluid quantities appearing in Eqs.~\eqref{eq:newton}--\eqref{eq:partheat} are evaluated at the position of each particle via trilinear interpolation. Further details can be found in \citep{Capecelatro2013}. 

\section{Results}
\label{sec:results} 
In this section, we summarize the results of the Euler--Lagrange simulations carried out using the setup and parameters discussed in Sec.~\ref{sec:problemStatement} and the computational framework laid out in Sec.~\ref{sec:mathematicalDescription}. We begin by reporting high level observations of both the cold-flow and thermal simulations, and then show profiles for the mean temperatures and quantify the thermal entrance length for each case. Finally, we propose scaling relations for the thermal entrance length corresponding to uncorrelated (uniform) particles and another that takes clustering into account.

\subsection{Flow visualization}
\begin{figure}[ht]
    \centering
    \includegraphics[width=\textwidth]{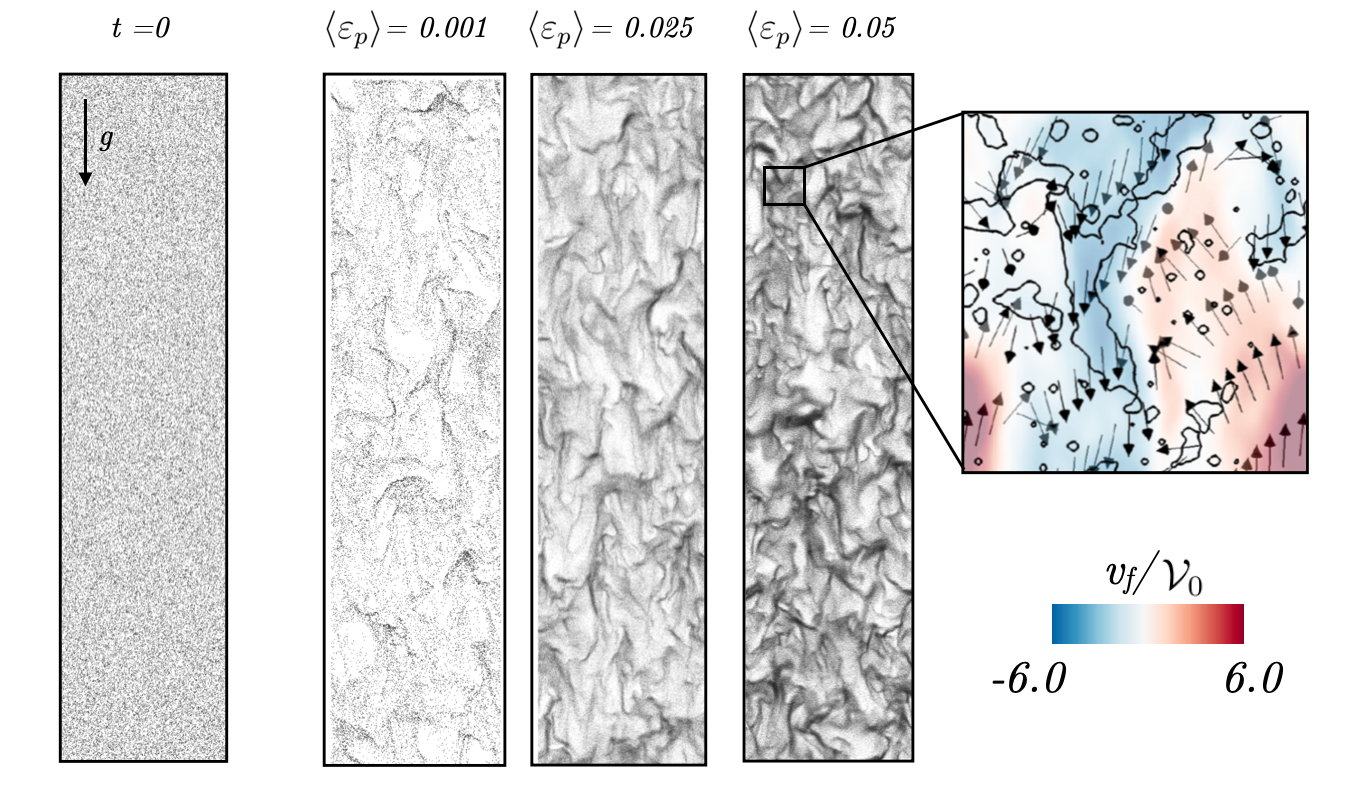}
    \caption{The isothermal simulations begin with an initially random distribution of particles (left) and evolve into a statistically stationary state characterized by clustering (middle 3 panels). These clusters generate and sustain turbulence in the fluid phase. Clusters entrain the fluid as they fall resulting in upflow in regions void of particles (right panel). }
    \label{fig:isothermal}
\end{figure}

Three cold-flow (isothermal) simulations are performed using the parameters summarized in Table.~\ref{tab:parameters1}. Instantaneous snapshots are shown in Fig.~\ref{fig:isothermal}. Beginning from an initially random distribution of particles, particles fall under gravity while the mean mass flow rate of the gas phase is held constant, allowing for a mean slip velocity between the phases to be established and a statistically stationary state to be reached after approximately $50\tau_p$. The degree of clustering is seen to vary as the volume fraction is increased. Dense suspensions of particles entrain the gas phase downward, resulting in so-called jet bypassing (high-speed upward flow in regions devoid of particles). At this point, the fluid-phase turbulent kinetic energy is produced by wakes behind clusters and shear layers at the edge of clusters, referred to as fully-developed cluster-induced turbulence (CIT)~\citep{capecelatro2014cit,capecelatro2015}.

\begin{figure}
    \centering
    \includegraphics[width = 0.8\textwidth]{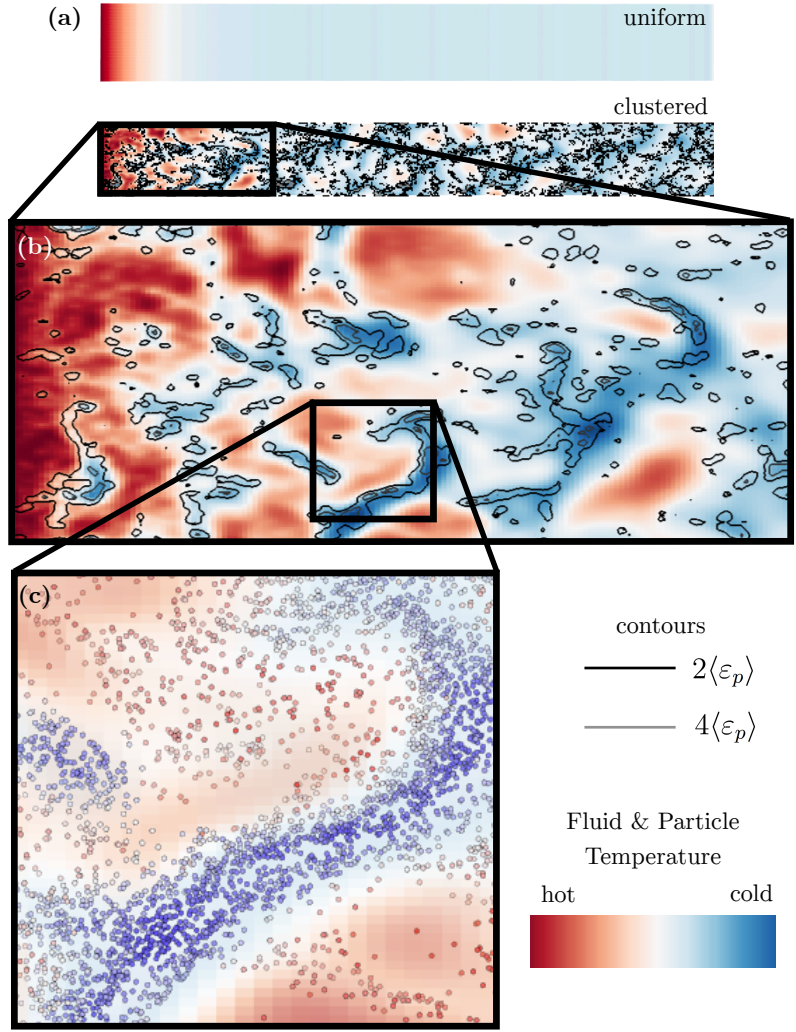}
    \caption{Hot (red) gas and cold (blue) particles are fed into a hot, quiescent thermal domain. From top to bottom: (a) When the particle phase is uncorrelated (uniformly distributed), the thermal entrance length is shorter as compared with a correlated (clustered) configuration of equal solid-phase volume fraction. (b) Clusters generate heterogeneity in the velocity (not shown) and temperature fields and (c) dilute regions of particles are heated rapidly, while denser clusters of cold particles persist further into the domain.  Images correspond to a instantaneous snapshots for $\langle \varepsilon_p \rangle = 0.001$, Pe $= 5$ and $\chi = 829$. A high-resolution video of this image can be found in the supplementary materials.}
    \label{fig:snapshot}
\end{figure}

In the thermal simulations, cool particles and hot gas from fully-developed CIT are injected into the thermal domain and heat transfer is enabled between the phases. As shown in Fig.~\ref{fig:snapshot}, the cool centers of dense clusters persist far into the domain and cool the surrounding fluid, while regions of dilute particles are heated more rapidly and have a minimal effect in cooling the fluid. This behavior is observed to be more dramatic for lower volume fraction and low \Peclet\; number. Not surprisingly, as the volume fraction is increased, the increase in mass loading of cold particles can more rapidly cool the surrounding gas, though hot spots still appear in regions devoid of particles. This behavior is shown in Fig.~\ref{fig:DevLengthCompare}.

\subsection{The thermal entrance length}
\label{subsec:effectClustering} 
To quantify the thermal entrance length, we extend the definition from single-phase pipe flow to the configuration under consideration. Since both phases relax to an equilibrium temperature, a nondimensional thermal entrance length, ${l}_{th}$, can be defined as the location after which the difference between the mean temperatures is within 5\% of the inlet temperature difference, or 
\begin{equation}
    l_{th} := \min\left(\hat{x} \in \vert\langle \theta_f \rangle - \langle \theta_p \rangle \vert \leq 0.05 \right), 
\end{equation}
where $\theta_{f/p}$ is the nondimensional temperature (for the fluid or particle phase) given by
\begin{equation}
    \theta_{f/p}=\frac{T_{f/p}-T_{p,0}}{T_{f,0}-T_{p,0}}
\end{equation}
and $\hat{x}=d/d_p$ is the nondimensional streamwise position.
\begin{figure}[ht]
    \centering
    \begin{tabular}{c c c}
    \includegraphics[height=0.25\textwidth]{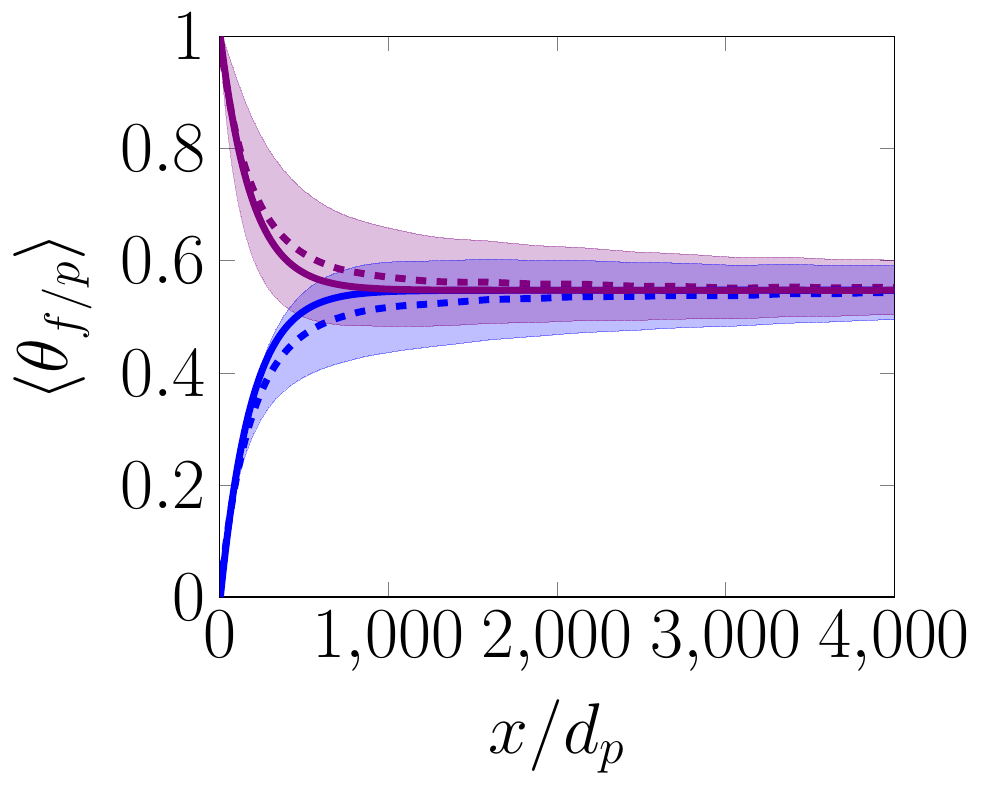} &
    \includegraphics[height=0.25\textwidth]{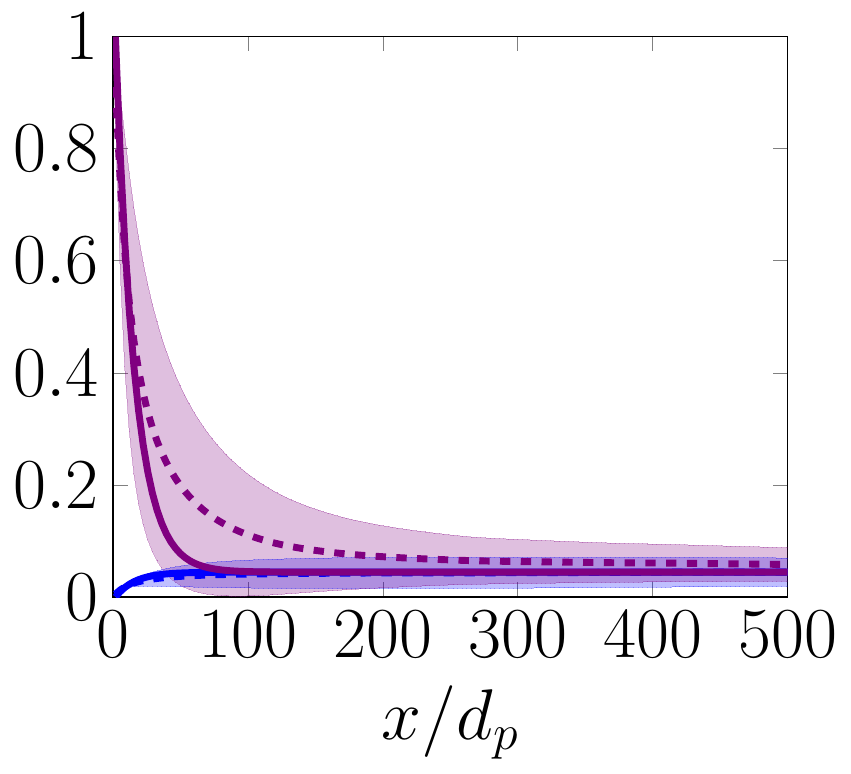}&
    \includegraphics[height=0.25\textwidth]{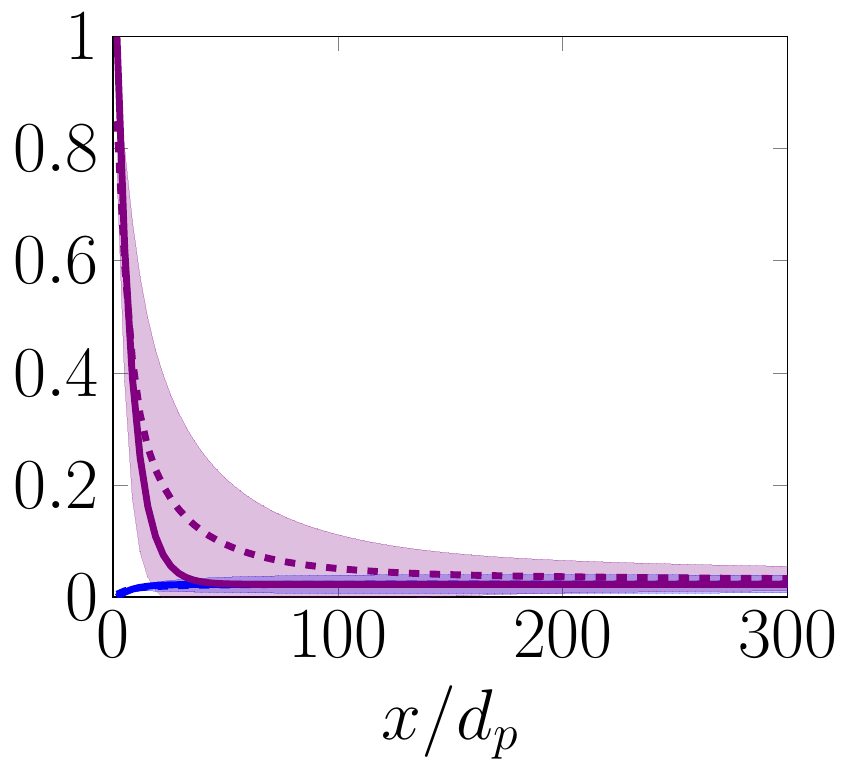}\\
    \multicolumn{3}{c}{\hspace{1em} \includegraphics[width=0.9\textwidth]{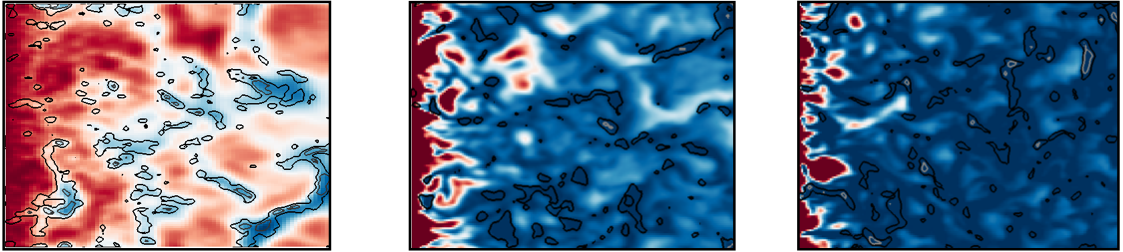}}
    \end{tabular}
    \caption{Temperatures are compared for the three volume fractions under study ($\langle \varepsilon_p \rangle = 0.001, 0.0255$ and $0.05$, from left to right) and $(\text{Pe},\chi) = (5,829)$.  The top row shows the mean temperature profiles for a uniform distribution of particles (\protect \uncorrf, \protect \uncorrp) and the Euler--Lagrange simulations (\protect \ELf, \protect \ELp), where the shaded regions represent the $\pm \sigma$, where $\sigma$ is the standard deviation. The bottom row shows the fluid temperature in the region between the inlet and $\hat{x} = l_{th}$. Red corresponds to high temperature and blue to low. The contours denote $\varepsilon_p = 2 \langle \varepsilon_p \rangle$. }
    \label{fig:DevLengthCompare}
\end{figure}

\begin{figure}[ht]
    \centering
    \includegraphics[width = 0.75\textwidth]{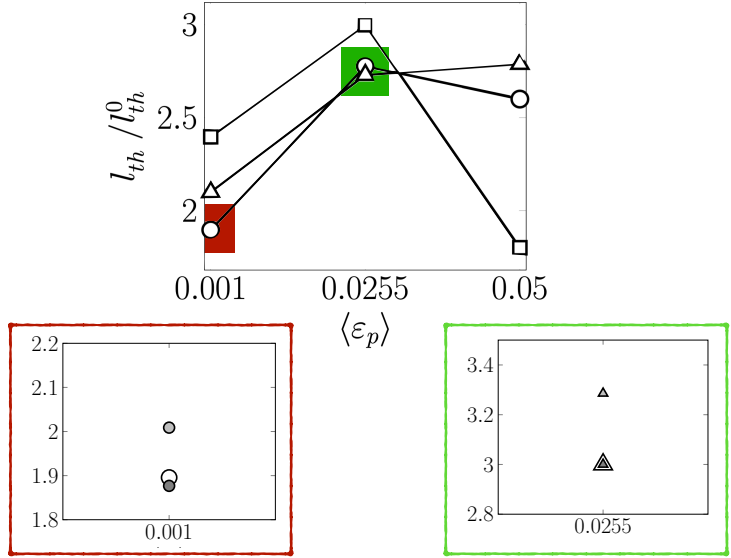}
    \caption{The entrance length normalized by the entrance length for a uniform distribution of particles of equivalent volume fraction (top). Here, \Peclet\; numbers 1, 5 and 7 are denoted by squares, circles and triangles, respectively. The inset bottom two plots examine the effect of $\chi$, where white, light gray and dark gray denote $\chi = (829, 909, 2270)$. }
    \label{fig:DevLengthSummary}
\end{figure}

The thermal entrance lengths for the clustered, Euler--Lagrange results are compared with the development lengths for a uniform distribution of particles of equivalent mean volume fraction, $l_{th}^0$ (see Figs.~\ref{fig:DevLengthCompare} and ~\ref{fig:DevLengthSummary}). By making this comparison, the effect of heterogeneity on interphase heat exchange can be isolated. Further, since the effect of clustering appears as a subgrid scale term in coarse-grained models~\citep{Rauchenzauner2020, Guo2019}, the ratio of these quantities highlights the errors associated with neglecting these contributions. 
 
 For the cases considered, it is observed that the thermal equilibrium temperature is lowered with increasing volume fraction, owing to the increased mass loading of cool particles as previously discussed. Additionally, the thermal entrance length decreases with increasing volume fraction, but in all cases, the presence of clusters acts to increase the thermal entrance length as compared with an uncorrelated distribution of particles. This can be seen in greater detail in Fig.~\ref{fig:snapshot} and is primarily a consequence of the reduced contact with a hot fluid phase, making clustered particles less effective at cooling the surrounding fluid than lone particles. Finally, in Fig.~\ref{fig:DevLengthCompare}, the shaded regions represent $\pm 3$ standard deviations from the mean temperature. This variation in temperature is greater in the fluid phase as compared with the particle phase, and the overall variation in temperature reduces with increasing volume fraction. 

In Fig.~\ref{fig:DevLengthSummary}, the entrance length obtained from the simulations are normalized by $l_{th}^0$, as previously mentioned, and compared against volume fraction, Pe and $\chi$. Here, we observe that for all the configurations considered, the entrance length for clustered particles is between 2 and 3 times longer than a uniform distribution, but that this relationship is complexly related to volume fraction and \Peclet\; number, in particular. Notably, the development length increases \emph{non-monotonically} with particle volume fraction for moderate and high \Peclet\; numbers, which is likely explained by similar behavior observed in the normalized standard deviation of the volume fraction, $\sqrt{\langle\varepsilon^{\prime 2}_p \rangle}/\langle \varepsilon_p \rangle$, a measure of the degree of clustering \cite[see][]{Beetham2021}. Finally, we observe that the ratio of heat capacities, $\chi$, has a relatively minimal effect on the thermal entrance length for clustered flows as compared with the entrance length for unclustered flows, $l_{th}^0$. This is shown in the inset panels in Fig.~\ref{fig:DevLengthSummary} and indicates that the thermal entrance length for clustered flows does not change significantly as compared $l_{th}^0$. This implies that models for capturing heterogeneous behavior should depend only on $\langle \varepsilon_p \rangle$ and the \Peclet\; number. 

Finally, we compute $l_{th}^0$ over a range of volume fraction, \Peclet\;, Reynolds and Prandtl numbers, and identify the following scaling relation for the thermal entrance length for a uniform distributions of particles, 
\begin{equation}
\label{eq:lth-uni}
    l_{th}^{0} = 0.108\; \Reyn_{\text{bulk}} \; \Pran \; \langle \varepsilon_p \rangle^{-1}.
\end{equation}
In this expression, the existence of the particles augments the single-phase expression~\eqref{eq:lth-single} by a factor of $0.216 \langle \varepsilon_p \rangle^{-1}$. This quantifies the observation that the entrance length increases with decreasing solids volume fraction and increasing Reynolds number. Here, this increase is nearly exponential with respect to volume fraction and linear with respect to Re$_\text{bulk}$. The l-2 norm of the error of the scaling relation for Re$_\text{bulk}$ $\in [0.2, 22]$ and $\langle \varepsilon_p \rangle \in [0.001,0.5]$ is 0.02.  
A similar scaling relation can be formulated for clustered flows, given as 
\begin{equation}\label{eq:lth}
l_{th} = 0.64\frac{\sqrt{\langle \varepsilon_p^{\prime 2} \rangle}}{\langle \varepsilon_p \rangle}\left(0.1 \frac{\text{Re}_{\text{bulk}}}{\langle \varepsilon_p \rangle} + 0.02 \;\text{Re}_{\text{bulk}}^3 \right) + 0.108\; \text{Re}_{\text{bulk}}\;\text{Pr}\;\langle{\varepsilon_p}\rangle^{-1},
\end{equation}
where the variance in volume fraction is informed by a modified version of the model developed by \citet{Issangya2000} given as
\begin{equation}\label{eq:epsvar}
\sqrt{\langle \varepsilon_p^{\prime 2} \rangle} = 1.48\langle\varepsilon_p \rangle\left(0.55-\langle \varepsilon_p\rangle \right).
\end{equation}
In this expression, the first coefficient differs from the original model of \citet{Issangya2000}, 1.584, to fit our data more accurately. The model \eqref{eq:lth} returns the scaling for an unclustered configuration, $l_{th}^0$, when particles are uncorrelated (i.e., $\langle \varepsilon_p^{\prime 2} \rangle=0$). This scaling relation has a normalized l-2 error norm of 0.04 for the data considered in this study. 

In the following section, we quantify the terms responsible for the complex behavior we observed in the thermal entrance length and propose closure to predict it over a range of conditions. 

\section{Modeling} 
\label{sec:modeling} 
In the previous section, we demonstrated that the thermal entrance length for clustered flows varies significantly from their uniform flow counterparts. Additionally, we observed that these differences depend complexly on the mean particle volume fraction as well as the \Peclet\; number. To quantify the effect that correlated phases has on this phenomenon, we first derive the one-dimensional, two-fluid heat equations. Next, we evaluate the contributions of each of the terms appearing in the thermal balance and propose models for the dominant unclosed term.

\subsection{One-dimensional heat equations} 
\label{subsec:PA}
For the configurations under consideration, the flow is statistically stationary in time, statistically homogeneous in the spanwise directions and thermally evolving in the streamwise direction. This implies that all quantities of interest are one-dimensional in $x$. To formulate the associated 1D heat equations, we first nondimensionalize the heat equation, then conduct volume fraction-weighted (phase) averaging. Nondimensionalization is carried out by selecting the particle diameter, $d_p$, as a characteristic length scale and the inlet bulk velocity, $u_{\rm bulk}$, as a characteristic velocity. Details on both of these derivations can be found in \ref{app:nonndim} and \ref{app:phaseaverage}. 

Beginning with the fluid-phase heat equation \eqref{eq:fluidheat}, the nondimensional fluid temperature equation is given by
 \begin{equation}
{\rm Pe} \frac{\partial }{\partial \hat{t}} \left( \varepsilon_f  \theta_f \right) + {\rm Pe} \frac{\partial}{\partial \hat{\bm{x}}} \left( \varepsilon_f \hat{\bm{u}}_f \theta_f \right) = \varepsilon_f\frac{\partial^2\theta_f }{\partial \hat{\bm{x}}^2}    - 6  {\rm Nu}\varepsilon_p \left( \theta_f - \theta_p \right),
\end{equation}
where $\hat{x} = x/d_p$, $\hat{u}_f = u_f/u_{\text{bulk}}$ and $\hat{t} = t/(d_p/u_{\rm bulk})$.
The particle-phase heat equation~\eqref{eq:partheat} can be similarly nondimensionalized. First, the particle phase heat equation is rewritten in the Eulerian sense by conducting a change of frame from the Lagrangian particle heat equation and projecting it to the Eulerian grid (see \ref{app:nonndim}). Then, in the same manner as the fluid heat equation, nondimensionalization yields 
\begin{equation}
     \chi \; {\rm Pe} \frac{\rho_p}{\rho_f}\left \lbrack \frac{{\partial} \left(\varepsilon_p \theta_p \right)}{{\partial} \hat{t}} + \frac{\partial \left( \varepsilon_p \hat{\bm{u}}_p  \theta_p \right)}{\partial \hat{\bm{x}}} \right \rbrack = \varepsilon_p \frac{{\partial}^2 \theta_f}{{\partial} \hat{\bm{x}}^2} + 6 {\rm Nu} \varepsilon_p \left (\theta_f - \theta_p \right),
\end{equation}
where, $\chi$ is the ratio of heat capacities. 

Next, Reynolds averages (denoted with angled brackets) in time and the spanwise directions are applied in order to treat these expressions as statistically one-dimensional. In doing so, the time derivative is null and due to periodicity in the $y-$ and $x-$ directions, gradients and divergence operators reduce to full derivatives with respect to $x$. This yields 
\begin{equation}
    {\rm Pe} \frac{{\rm d} \langle \varepsilon_f \hat{{u}}_f \theta_f \rangle }{{\rm d} \hat{{x}}}  = \frac{{\rm d}^2 \langle \varepsilon_f\theta_f \rangle }{{\rm d} \hat{{x}}^2}    - \langle 6  {\rm Nu}\varepsilon_p \left( \theta_f - \theta_p  \right)\rangle \label{eq:1Dfluid}
\end{equation}
and
\begin{equation}
    \chi \; {\rm Pe} \frac{\rho_p}{\rho_f} \frac{\deriv \langle  \varepsilon_p \hat{u}_p \theta_p  \rangle}{\deriv \hat{x}} =  \frac{{\rm d}^2 \langle \varepsilon_p \theta_p \rangle }{{\rm d} \hat{{x}}^2} + \langle 6 {\rm Nu} \varepsilon_p \left (\theta_f - \theta_p \right)\rangle. \label{eq:1Dpart} 
\end{equation}
As will be discussed later, the diffusion terms are found to have a minimal contribution to the thermal balance in both phases, but they are included here since the \Peclet\; numbers are $\mathcal{O}(1)$. In cases of very large \Peclet\; number, however, the diffusion term can be eliminated \emph{a priori} due to the factor $1/{\rm Pe}$ that multiplies it.

Due to the presence of the volume fraction on all terms, a phase average  defined as $\langle{(\cdot)}\rangle_f = \langle \varepsilon_f (\cdot) \rangle / \langle \varepsilon_f \rangle$ and $\langle{(\cdot)}\rangle_p = \langle \varepsilon_p (\cdot) \rangle / \langle \varepsilon_p \rangle$ (as described in \citet{fox2014}) is convenient to invoke. This process substantially reduces the number of terms present as compared with strict Reynolds averaging. 

In these expressions, angled brackets without a subscript, $\langle (\cdot ) \rangle$, denote a Reynolds average in time and the cross stream directions (i.e., $y$ and $z$). Here, fluctuations from Reynolds averages are denoted as $(\cdot)^{\prime}$ and fluctuations from the particle and fluid phase averages are denoted as $(\cdot)^{\prime \prime}$ and $(\cdot)^{\prime \prime \prime}$, respectively. This yields a coupled system of two-fluid equations that may be utilized to model macroscopic heat transfer (noting that the solution of the momentum equations is trivial for the configuration under consideration), given as

\begin{align}
     &  \langle{\hat{u}_f}\rangle_f \frac{\deriv \langle{\theta_f}\rangle_f}{ \deriv \hat{x}}  - \frac{1}{ {\rm Pe} } \frac{\deriv^2 \langle \theta_f \rangle_f}{\deriv \hat{x}^2} = - \underbrace{\frac{\deriv}{\deriv \hat{x}} \langle{\hat{u}_f^{\prime \prime \prime} \theta_f^{\prime \prime \prime}}\rangle_f}_{\rm Term \; 1} \nonumber \\
     - & \frac{6 \langle \varepsilon_p \rangle}{ {\rm Pe} \; \langle \varepsilon_f \rangle} \Big[ \underbrace{\langle {\rm Nu} \rangle_p \left(\langle \theta_f \rangle_f - \langle \theta_p \rangle_p \right)}_{\rm Term \; 2} + \underbrace{\langle {\rm Nu} \rangle_p \langle \theta_f^{\prime \prime \prime} \rangle_p}_{\rm Term \; 3} +  \underbrace{\langle {\rm Nu}^{\prime \prime} \theta_f^{\prime \prime} \rangle_p}_{\rm Term \; 4} - \underbrace{\langle {\rm Nu}^{\prime \prime} \theta_p^{\prime \prime} \rangle_p}_{\rm Term \; 5}  \Big] \label{eq:PATf}
\end{align}
and
\begin{align}
     & \langle \hat{u}_p \rangle_p \frac{\deriv \langle \theta_p \rangle_p}{\deriv \hat{x}} - \frac{\rho_f}{\rho_p\chi \; {\rm Pe} } \frac{\deriv^2 \langle \theta_p \rangle_p }{\deriv \hat{x}^2}= -\underbrace{\frac{\deriv}{\deriv \hat{x}} \langle \hat{u}_p^{\prime \prime} \theta_p^{\prime \prime} \rangle_p}_{\rm Term \; 6}   \nonumber \\
&+\frac{6 \rho_f }{ \rho_p \chi \; {\rm Pe}} \Big[ \underbrace{\langle {\rm Nu} \rangle_p \left(\langle \theta_f \rangle_f - \langle \theta_p \rangle_p \right)}_{\rm Term \; 2} + \underbrace{\langle {\rm Nu} \rangle_p \langle \theta_f^{\prime \prime \prime} \rangle_p}_{\rm Term \; 3} +  \underbrace{\langle Nu^{\prime \prime} \theta_f^{\prime \prime} \rangle_p}_{\rm Term \; 4} - \underbrace{\langle {\rm Nu}^{\prime \prime} \theta_p^{\prime \prime} \rangle_p}_{\rm Term \; 5}  \Big]. \label{eq:PATp}
\end{align}

The terms in these expressions can be categorized as purely fluid, purely particle and mixed. The terms on the left-hand side of the particle and fluid equations represent convection and diffusion and are purely fluid and purely particle, respectively. Terms 1 and 6 are scalar fluxes, which are unclosed and traditionally modelled by classical gradient diffusion models (e.g., the Boussinesq approximation). While these methods are successful in single-phase flows, they have been shown to fall short of being predictive in the context of highly anisotropic, multi-phase flows~\citep{capecelatro2018biomass}. Finally, the interphase heat exchange terms (Terms 1--5), are the same across the fluid and particle phase descriptions, with the exception of a constant factor of $\rho_f/(\rho_p\chi)$ that appears in the particle phase equation and a factor of $\langle \varepsilon_p \rangle/\langle \varepsilon_f\rangle$ in the fluid phase. For brevity, these two factors are referred to as $C_1$ and $C_2$ henceforth. Of the interphase heat exchange terms, only Term 2 is a function of solution variables ($\langle \theta_f \rangle_f, \langle \theta_p \rangle_p$) and are therefore closed. Term 3 includes a covariance between volume fraction and fluid temperature fluctuations, $\langle \theta_f^{\prime \prime \prime}\rangle_p $, which has been shown to be the main contributor to hindering heat transfer in temporally evolving, homogeneous systems \citep{Guo2019}. Terms 4 and 5 are cross correlations between phase temperature and Nusselt number. 


In the absence of clustering (e.g., no correlation between temperature and volume fraction), the only terms that remain are convection, diffusion and Term 2. Following from the definition of the phase average, $\langle \theta_f \rangle_f = \langle \theta_f \rangle + \langle \varepsilon_f^{\prime} \theta_f^{\prime} \rangle/\langle \varepsilon_f \rangle$, thus, in an uncorrelated, homogeneous system, $\langle \theta_f \rangle_f$ is equivalent to $\langle \theta_f \rangle$ and $\langle \theta_p \rangle_p = \langle \theta_p \rangle$) since cross correlations are null. 

\subsection{Thermal budget} 
\label{subsec:effectBalance} 

\begin{figure}
    \centering
    \begin{tabular}{c c c}
    \includegraphics[height = 0.25\textwidth]{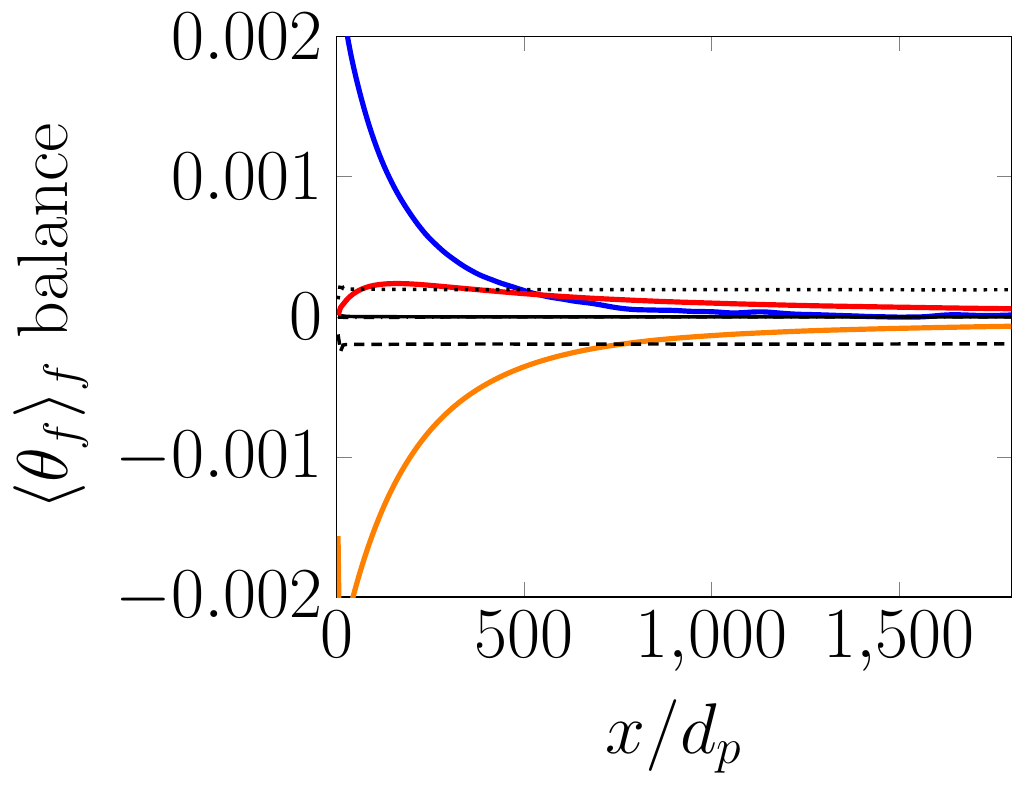} &
     \includegraphics[height = 0.25\textwidth]{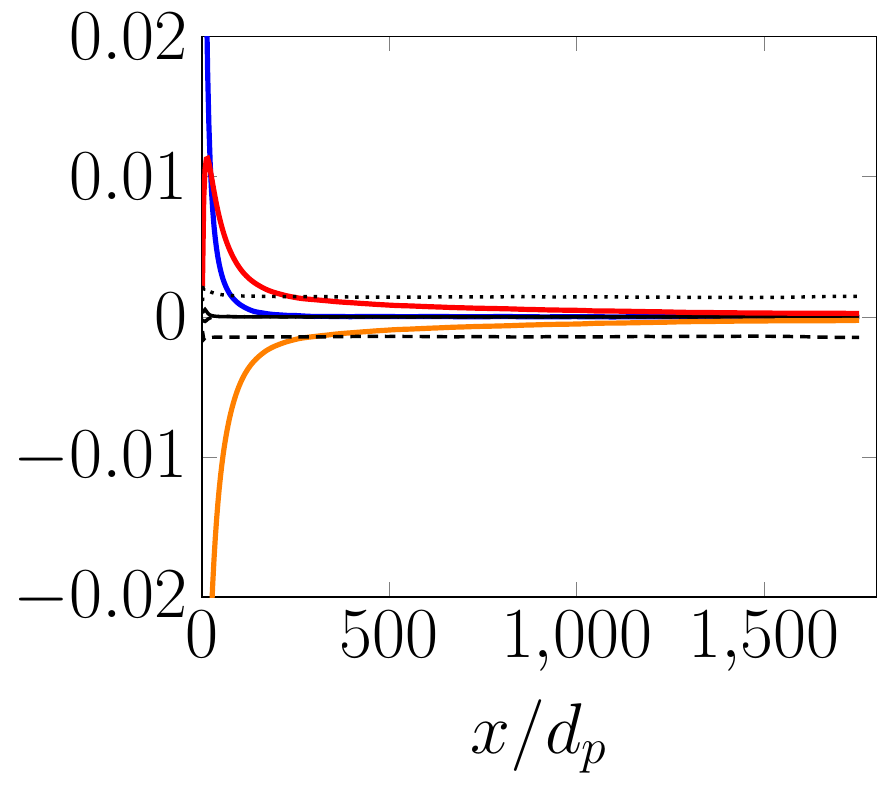} &
      \includegraphics[height = 0.25\textwidth]{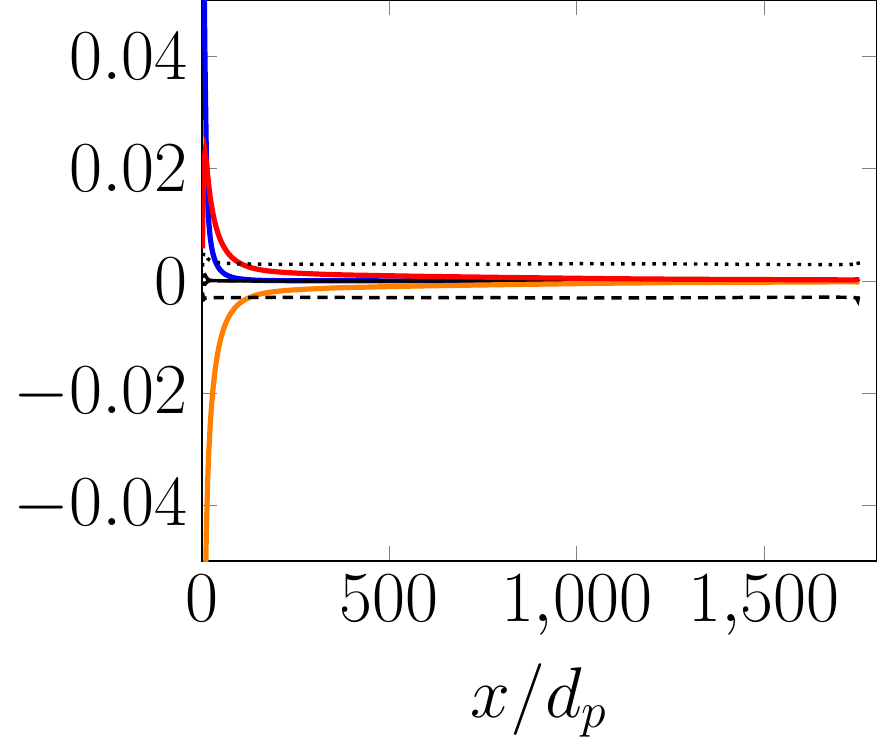}
    \end{tabular}
    \caption{Balance of terms contributing to the phase averaged fluid temperature as given in Eq.~\eqref{eq:PATf} for $Pe = 5$, $\chi = 829$. In each of the three volume fractions (0.001, 0.0255 and 0.05 from left to right), three dominate the thermal behavior: Convection (blue), Term 2 (orange) and Term 3 (red).}
    \label{fig:balance}
\end{figure}

To guide our modeling efforts, we now evaluate which of the terms appearing in \eqref{eq:PATf} and \eqref{eq:PATp} have leading order effects. As such, the balance of these terms is shown in Fig.~\ref{fig:balance} for the illustrative case of Pe $= 5$ and $\langle \varepsilon_p \rangle = 0.0255$. This demonstrates that for the configuration under consideration, thermal behavior is dominated by convection, Term 2 and Term 3. Of these terms, only the fluid phase temperature fluctuations as seen by the particles, $\langle \theta_f^{\prime \prime \prime} \rangle_p$ (defined as the `drift temperature' in \citet{Rauchenzauner2020}), requires modeling and is equivalent across both phases. 

\subsection{Closure of the drift temperature}


In this section, we propose a closure model for $\langle \theta^{\prime \prime \prime}_f \rangle_p$ and equivalently, $\langle \varepsilon_p^{\prime} \theta_f^{\prime} \rangle$ (see~\ref{sec:RAvsPA}). As previously mentioned, and by definition of the phase average, the phase averaged temperatures are comprised of the Reynolds averaged temperature plus the cross correlation between volume fraction and temperature (i.e., $\langle \theta_f \rangle_f = \langle \theta_f \rangle + \langle \varepsilon_f^\prime \theta_f^{\prime}\rangle/\langle \varepsilon_f \rangle$). Because of this, specifying boundary conditions for the heat equations in terms of phase-averaged quantities cannot be done \emph{a priori} without an additional closure for these contributions. Rather than providing additional closures (one each for $\langle \varepsilon_p^{\prime} \theta_p^{\prime} \rangle$ and $\langle \varepsilon_f^{\prime} \theta_f^{\prime} \rangle$), we note that for the configuration under study, the cross correlations are constant with respect to the streamwise direction and only shift the temperature solution by this amount. In other words, the thermal entrance length is equivalent when considering either the phase averaged \emph{or} Reynolds averaged temperatures, but the Reynolds averaged formulation does not require special treatment for the boundary conditions, as they are specified as the same for clustered and unclustered flows. (See \ref{sec:RAvsPA} for more details). 

\begin{figure}[ht]
    \centering
    \includegraphics[width=0.45\textwidth]{figures/DevLengthCase4_model.pdf} 
    \includegraphics[width=0.45\textwidth]{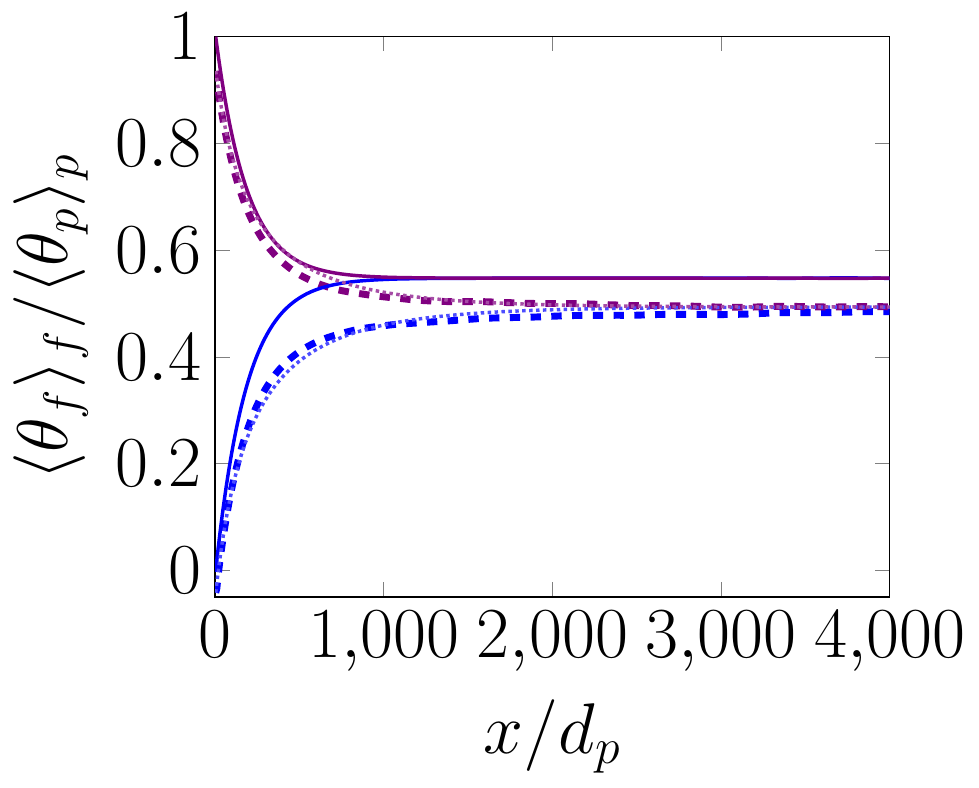}
    \caption{The learned model (shown as (\protect \corrf, \protect \corrp) and described in Eq.~\eqref{eq:model}) demonstrates improved prediction of thermal entrance length as compared to the Euler--Lagrange results (\protect \ELf, \protect \ELp) in both the Reynolds-averaged (left) and the phase-averaged formulations (right). The forward solution using the assumption of uniformly distributed particles is shown as (\protect \uncorrf, \protect \uncorrp), and is the same in both plots since phase averaging and Reynolds averaging are equivalent when the phases are uncorrelated. }
    \label{fig:RAvsPA}
\end{figure}

Due to the equivalency of $\langle \varepsilon_p^{\prime} \theta_f^{\prime} \rangle$ and $\langle \theta^{\prime \prime \prime}_f \rangle_p$, the model proposed herein is suitable for use in simulations for which the solution variables are phase-averaged \emph{or} Reynolds averaged (demonstrated in Fig.~\ref{fig:RAvsPA} where the proposed model detailed in Sec.~\ref{sec:modeling} is used in forward solutions of both sets of state variables). Thus implying it is appropriate for use in a general two-fluid solver in which the hydrodynamics and thermodynamics evolve simultaneously. Of course, in this situation, additional closures are required for the fluid and particle momentum equations in order to capture cross correlations.

We begin from the simplified, Reynolds averaged equations, 
\begin{equation}
  \langle \hat{u}_f \rangle \frac{\deriv \langle \theta_f \rangle}{\deriv \hat{x}}= - \frac{6 \langle \varepsilon_p \rangle \widetilde{Nu}}{\text{Pe}\langle \varepsilon_f \rangle} \left \lbrack \langle \theta_f \rangle - \langle \theta_p \rangle +\frac{\langle \varepsilon_p^{\prime} \theta_f^{\prime}\rangle}{\langle \varepsilon_p \rangle}
    \right \rbrack,
\end{equation} 
where $\widetilde{Nu}$ denotes the Nusslet number computed using the correlation proposed by \citet{Sun2019} and mean quantities as arguments. As detailed in~\ref{sec:RAvsPA}, all of the unclosed Reynolds averaged terms are null, except for the cross correlation between particle volume fraction and the fluid-phase temperature fluctuations arising from Term 3, as was observed for temporally evolving gas-particle flows from \citet{Guo2019}. 

This result points to the fact that cross-correlations between volume fraction and temperature shift the phase averaged temperature from the Reynolds averaged temperature (e.g., $\langle \theta_f (\hat{x}) \rangle_f = \langle \theta_f(\hat{x}) \rangle + \langle \varepsilon_f^{\prime} \theta_f^{\prime} \rangle $), where in these configurations the cross correlations are constant with respect to $\hat{x}$. 

In formulating the closure for $\langle\varepsilon_p^{\prime}\theta_f^{\prime}\rangle$, we observe that all configurations considered in this work satisfy the following scaling relation
\begin{equation} 
\frac{1}{\left(\langle \theta_f \rangle - \langle \theta_p \rangle \right)}\left(\frac{\deriv \langle \theta_f \rangle}{\deriv \hat{x}} + C_1\left(\langle \theta_f \rangle - \langle \theta_p \rangle \right)\right) =  b \left(\langle \theta_f \rangle - \langle \theta_p \rangle + 1\right)
\end{equation} 
where $b$ is a constant coefficient, which may depend upon $\langle \varepsilon_p \rangle$, $\chi$ and Pe. Owing to this relation, we impose that the the proposed model is of the form
\begin{equation}
    \frac{\langle \varepsilon_p^{\prime} \theta_f^{\prime} \rangle}{\langle \varepsilon_p\rangle} = -\frac{b}{C_1}\left( \langle \theta_f \rangle - \langle \theta_p \rangle \right)\left( \langle \theta_f \rangle - \langle \theta_p \rangle + 1)\right)
\end{equation}
\begin{figure}[ht]
    \centering
    \begin{tabular}{c c c}
    \multicolumn{3}{c}{$(\langle \varepsilon_p \rangle, \text{Pe}, \chi) $}\\ [0.5em]
    \hspace{2em} (0.001, 5, 829) & \hspace{2em}(0.0255, 7, 829) &  \hspace{2em}(0.05, 5, 829)    \\ [0.5em]
    \includegraphics[height = 0.25\textwidth]{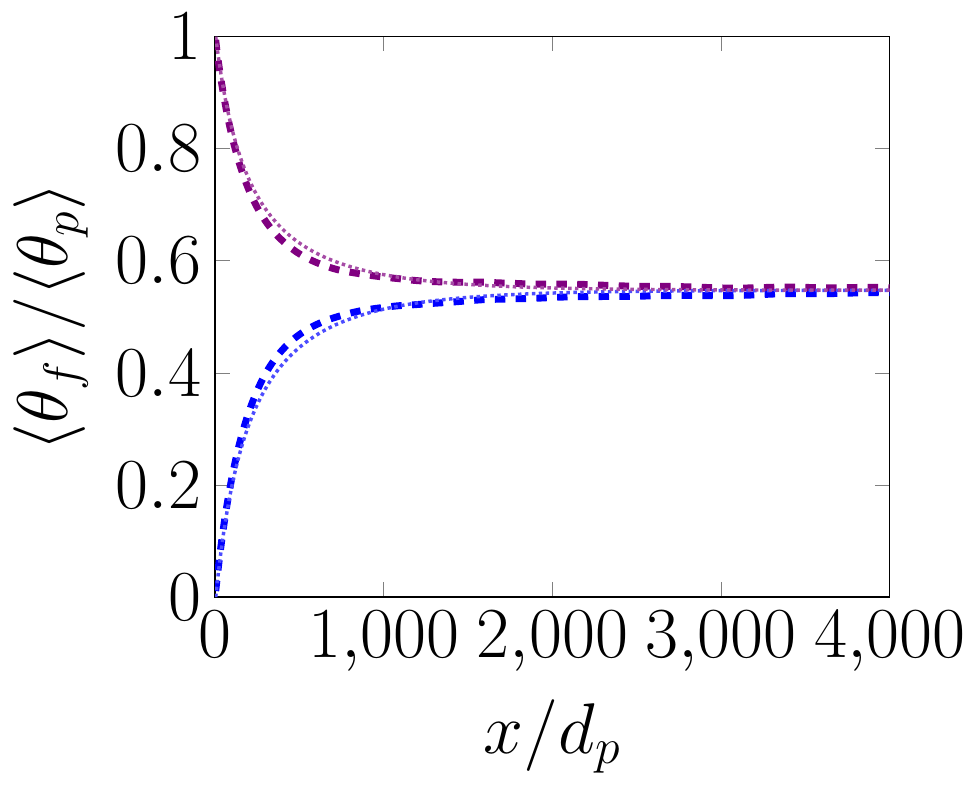} & 
      \includegraphics[height = 0.25\textwidth]{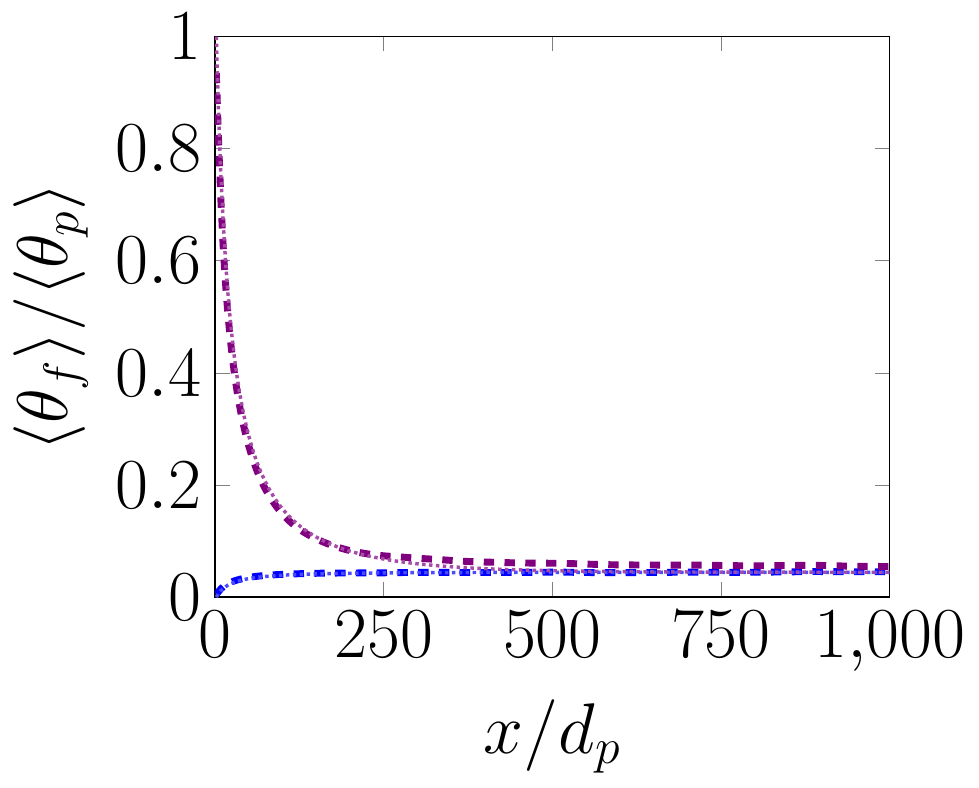} &
    \includegraphics[height = 0.25\textwidth]{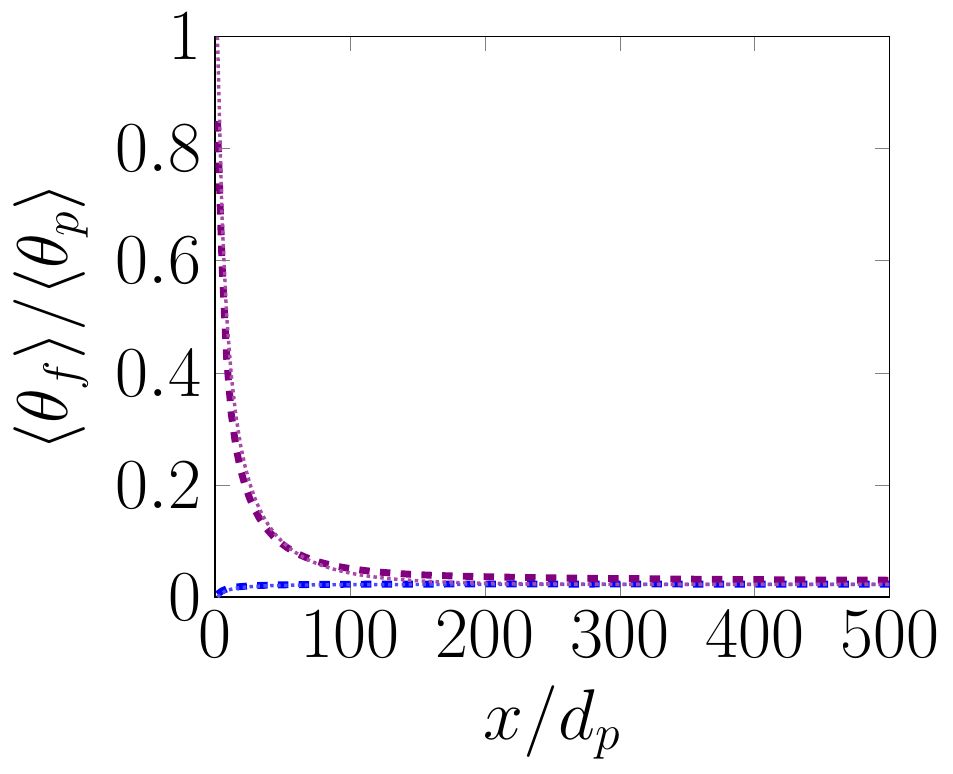} \\ [1em]
    \multicolumn{3}{c}{$(\langle \varepsilon_p \rangle, \text{Pe}, \chi) $}\\ [0.5 em]
    \hspace{2em} (0.001, 5, 909) &   \hspace{2em}(0.001, 5, 2270) &   \hspace{2em}(0.0255, 1, 909) \\ [0.5em]
    \includegraphics[height = 0.25\textwidth]{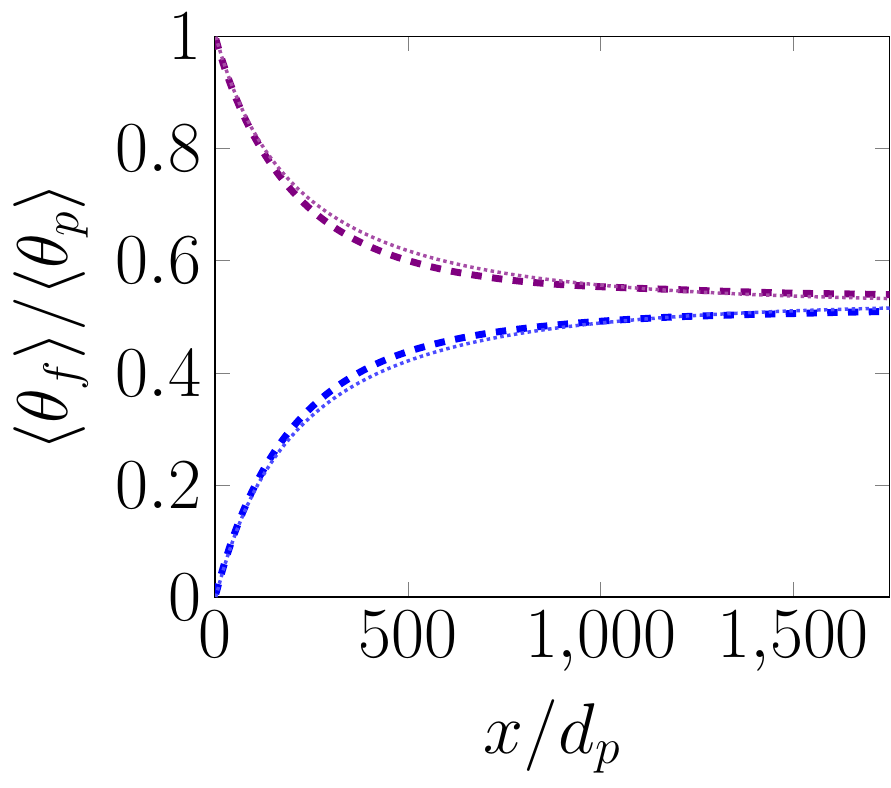} & 
    \includegraphics[height = 0.25\textwidth]{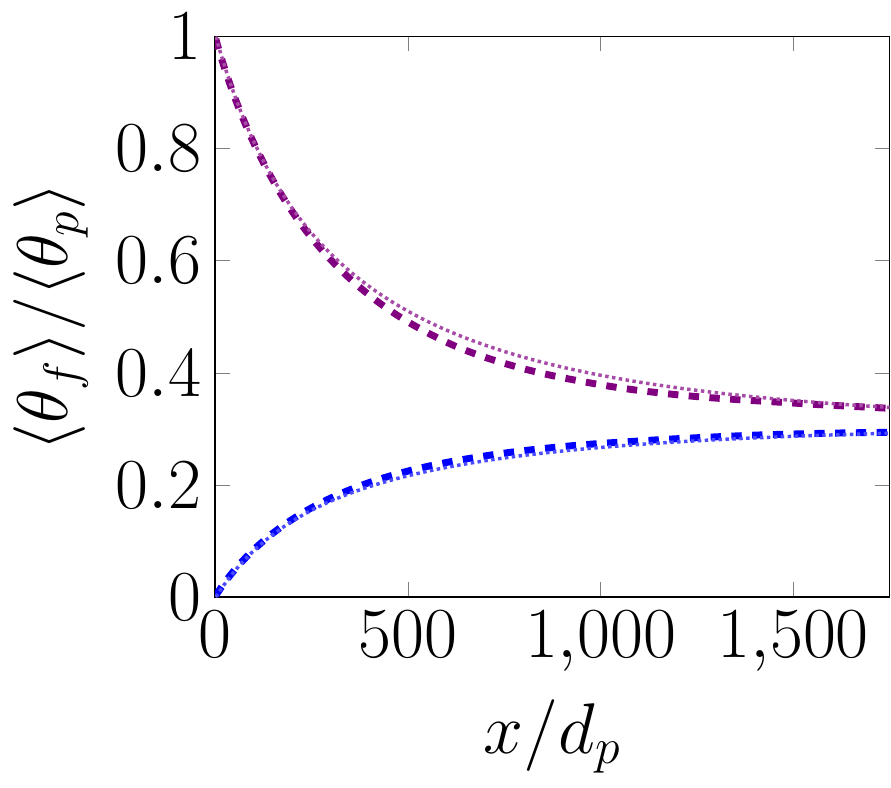} & 
    \includegraphics[height = 0.25\textwidth]{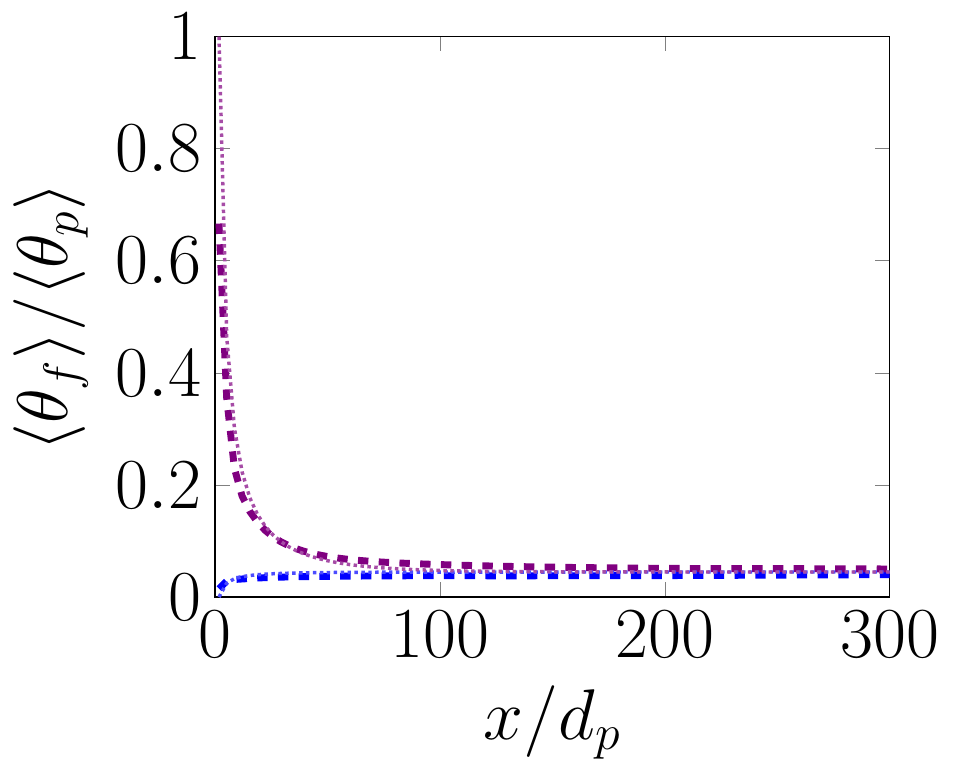} \\
    \end{tabular}
    \caption{Three example instances each of model performance (\protect \corrf, \protect \corrp) on training data (top row) and testing data (bottom row), as compared to the mean quantities  from Eulerian--Lagrangian data (\protect \ELf, \protect \ELp).}
    \label{fig:models}
\end{figure}
and the system of equations is given as
\begin{align}
\frac{\deriv \langle \theta_f \rangle}{\deriv \hat{x}} &= -C_1\left( \langle \theta_f \rangle - \langle \theta_p \rangle \right) + b \left( \langle \theta_f \rangle - \langle \theta_p \rangle \right)\left \lbrack \left( \langle \theta_f \rangle - \langle \theta_p \rangle \right) + 1\right \rbrack \\ 
\frac{\deriv \langle \theta_p \rangle}{\deriv \hat{x}} &= C_2\left( \langle \theta_f \rangle - \langle \theta_p \rangle \right) - \frac{ C_2}{C_1} b \left( \langle \theta_f \rangle - \langle \theta_p \rangle \right)\left \lbrack \left( \langle \theta_f \rangle - \langle \theta_p \rangle \right) + 1\right \rbrack. 
\end{align}

The open-source, Gene Expression Programming (GEP) MATLAB code of \citet{Searson2009}, is leveraged to learn the dependence of $b$ on operating parameters. The resultant model was selected from the models learned using a population size of 300, with 500 generations and a maximum number of genes per individual of 6. The GEP algorithm was provided with the value of $b$ and associated $\langle \varepsilon_p \rangle$ and $\Pec$ for each training case (all three volume fractions, all three \Peclet\; numbers and $\chi = 829$) and was permitted to evolve expressions from the following mathematical operations: multiplication/division, addition/subtraction, exponential/log, and square/cube. The resultant learned model for $b$ is given as
\begin{equation}
b = \left(1.16\ln(\langle \varepsilon_p \rangle) - 0.335 \text{Pe} + 5.85 \langle \varepsilon_p \rangle \text{Pe} + 19.7\right) \sqrt {\langle \varepsilon^{\prime 2}\rangle} \left(1 - e^{-\langle \varepsilon_p \rangle/\text{Pe}}\right), \label{eq:model}
\end{equation}
where the inclusion of the variance of volume fraction, $\sqrt{\langle \varepsilon^{ \prime 2} \rangle}$, in the expression for $b$ ensures proper asymptotic behavior in the limit of no clustering (i.e., Term 3 vanishes in the case of a uniform distribution of particles), which is modeled according to~\eqref{eq:epsvar}.

Figure~\ref{fig:models} highlights the forward solution of the proposed model and the forward solution for a uniform distribution of particles. Both are compared against the mean Euler--Lagrange data. The top and bottom rows show three representative training and test cases, respectively. Since the ratio of heat capacities was observed to have a minimal effect on entrance length as discussed in Sec.~\ref{sec:results}, perturbations in $\chi$ were reserved for the testing set. 

Additionally, the predicted entrance lengths for the uncorrelated forward solution and the forward solution with the proposed model for $b$ is summarized in Table~\ref{tab:Lth_summary} and compared against the Euler--Lagrange results. We find that using an assumption of uniformly distributed particles results in an under prediction of the thermal entrance length between 40 and 70\%. This highlights the importance of incorporating local particle heterogeneity in reduced-order (coarse grained) models. The proposed model demonstrates improved performance, predicting the thermal entrance length within 3.6\%, on average for the training dataset and within 8.6\% for the testing dataset. 

\begin{table}[ht]
    \centering
    \begin{tabular}{c c c @{\hskip 0.3in} r r r @{\hskip 0.3in} c c}
    \hline 
    \hline 
    $\chi$ & Pe & $\langle \varepsilon_p \rangle$ & $l_{th}$ & $l_{th}^0$ & $l_{th}^{\text{model}}$ & $\epsilon^{0}$ & $\epsilon^{model}$ \\
    \hline
    \multirow{9}{*}{840} & \multirow{3}{*}{1} & 0.001 &  258.9 & 108.0 & 214.3 & 0.6 & 0.026\\
                                           &  & 0.0255 & 36.0 & 12.0 & 32.5 & 0.7 & 0.095 \\
                                           &  & 0.005 & 15.4 & 8.6 & 15.4 & 0.4 & 0.00 \\ [0.5em]
                           & \multirow{3}{*}{5} & 0.001 &  1,059.5 & 558.9 & 1,114.4 & 0.5 & 0.045 \\ 
                                           &  & 0.0255 & 128.6 & 46.3 & 152.5 & 0.6 & 0.053 \\
                                           &  & 0.005 & 66.9 & 25.7 & 66.9 & 0.6 & 0.000 \\ [0.5em]
                         & \multirow{3}{*}{7} & 0.001 &   1,681.8 & 780.1 & 1,555.0 & 0.5 & 0.084 \\
                                              &  & 0.0255 & 173.2 & 63.4 & 210.9 & 0.6 & 0.020 \\
                                              &  & 0.05 & 90.9 & 32.6 & 94.3 & 0.6 & 0.000 \\ [0.5em]
      \hline
      \multicolumn{8}{c}{Testing Dataset}\\ 
      \hline
     \multirow{2}{*}{921} &{1} & 0.0255 &  39.4 & 1.2 & 32.5 & 0.7 & 0.174\\[0.25em]
      & {5} & 0.001 &  1,177.8 & 584.6 & 1,164.1 & 0.5 & 0.095\\ [0.5em]
     \multirow{2}{*}{2300} & {1} & 0.0255 &  36.0 & 1.2 & 32.5 & 0.7 & 0.014\\[0.25em]
     & {5} & 0.001 &  1,582.4 & 848.6 & 1,695.5 & 0.5 & 0.060\\
         \hline 
         \hline
    \end{tabular}
    \caption{Summary of thermal entrance lengths normalized by $d_p$ for clustered gas--solid flows and associated model errors. The learned model was trained on data for $\chi = 840$. Remaining cases were reserved for testing. On average, the entrance length predicted using an uncorrelated particle-phase assumption is under predictive by 58\%, while the prediction from the learned model predicts entrance length within 5.1\%, where the mean training and testing errors are 3.6\% and 8.6\%, respectively. }
    \label{tab:Lth_summary}
\end{table}

To make physical connections with the resultant model, it is helpful to introduce a new variable representing the temperature difference between the phases, $\langle \theta_\Delta \rangle = \langle \theta_f \rangle - \langle \theta_p \rangle$, and corresponding transport equation. This definition and some algebraic manipulation, yields the following system: 
\begin{align}
    \frac{\deriv \langle \theta_f \rangle}{\deriv \hat{x}} &= ({b-C_1})\langle \theta_\Delta \rangle \left( 1 - \frac{\langle \theta_\Delta \rangle }{(b-C_1)/b}  \right)\\
\frac{\deriv \langle \theta_p \rangle}{\deriv \hat{x}} &= \frac{C_2(C_1-b)}{C_1}\langle \theta_\Delta \rangle \left( 1- \frac{\langle \theta_\Delta \rangle}{(b-C_1)/b} \right)\\
\frac{\deriv \langle \theta_\Delta \rangle}{\deriv \hat{x}} &= \frac{\deriv \langle \theta_f \rangle}{\deriv \hat{x}} -  \frac{\deriv \langle \theta_p \rangle}{\deriv \hat{x}} \\
&= \left(-(C_1+C_2)+\frac{b(C_1-C_2)}{C_1}\right)\langle \theta_\Delta \rangle \left( 1- \frac{\langle \theta_\Delta \rangle}{(b-C_1)/b} \right), 
\end{align}
where we note that the equation for the mean temperature difference is of the same form as the logistic equation, i.e., $\deriv A/\deriv x = k A(1-A/L)$. In this sense, $L$ is frequently referred to as the limiting factor, or carrying capacity, of the system and $k$ is the growth rate. In the context of heat transfer for particle-laden flows, bifurcation points exist when either $\langle \theta_\Delta \rangle = 0$ or $\langle \theta_\Delta \rangle = L$. For this system and boundary conditions, the only physically relevant bifurcation occurs when the temperature difference is null. This point is a stable attractor, ensuring that all realizations with physical boundary conditions and parameters will relax to thermal equilibrium. 

\begin{figure}
    \centering
    \includegraphics[width=0.75\textwidth]{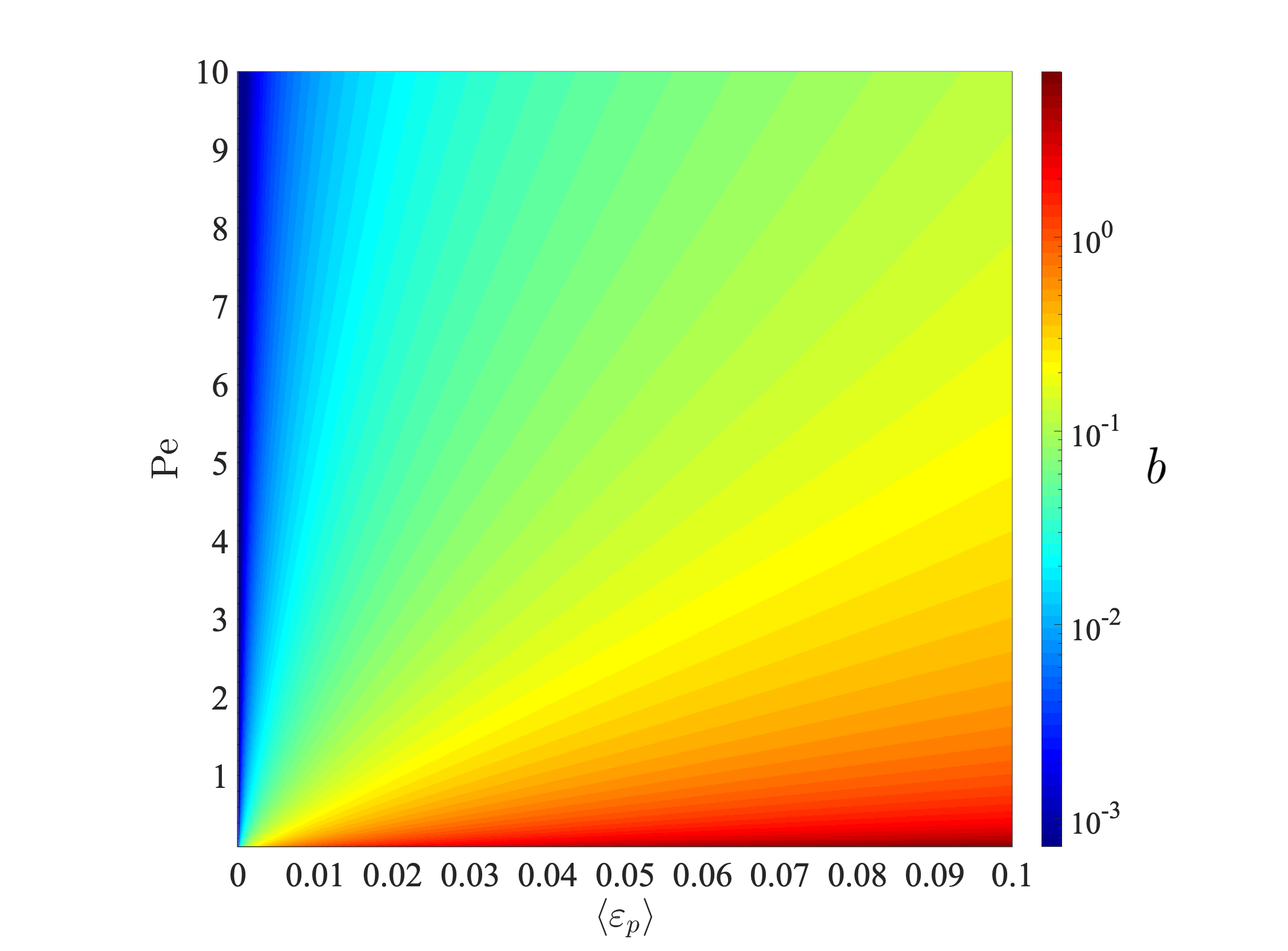}
    \caption{The modeled quantity, $b$ \eqref{eq:model}, shown with respect to $\langle \varepsilon_p \rangle$ and Pe.}
    \label{fig:b}
\end{figure}

Finally, the growth rate (which in this case is a negative value, indicating decay to equilibrium) is given as $-(C_1+C_2)+b(C_1-C_2)/C_1$. In the event of no clustering, the rate reduces to the uncorrelated growth rate $-(C_1+C_2)$, thus demonstrating that the presence of clusters impedes the rate at which the phases approach equilibrium. Written in this form, it can also be observed that the fluid- and particle-phase growth rates differ by a factor of $(-C_1/C_2)$, when clustering is present. Further, due to the dependence of $b$ on volume fraction and \Peclet\; number, the model quantifies the complex interplay of volume fraction and \Peclet\; number on thermal entrance length. This effect is visualized in Fig.~\ref{fig:b}, where we observe that for low \Peclet\; numbers, variations in volume fraction have a greater effect on the value of $b$. Similarly, at high volume fraction, changes in \Peclet\; number (particularly between 0 and 1) also result in large changes in $b$. Conversely, at high \Peclet\; number and as volume fraction approaches null, $b$ only changes slightly. Further, since $b$ implicates heat transfer impedance, one can expect longer thermal entrance lengths for lower volume fractions at higher \Peclet\; numbers and lower particle volume fractions.

\section{Conclusions}
\label{sec:conclusions} 
In this work, high resolution Euler--Lagrange simulations were leveraged to understand the effect of heterogeneity on the thermal entrance length. These computations enabled the quantification of the complex dependency of the entrance length on relevant simulation parameters, \Peclet\; number, volume fraction and ratio of phase heat capacities. In addition, we compared the thermal entrance length for clustered and uniform distributions of particles and found that clustering causes a 2 to 3 fold increase in $l_{th}$. To capture this effect, we propose a scaling relation for $l_{th}$ in Eq.~\eqref{eq:lth-uni} (for uniform distribution of particles) and Eq.~\eqref{eq:lth} (for clustered) that bares resemblance to scaling laws for the thermal entrance length of single-phase flows, but with an additional factor to account for the presence of particles. 

To identify the physics responsible for the change in thermal behavior of clustered flows, we derive the 1D two-fluid heat equations and evaluate which terms dominate. This analysis demonstrated that the delay in heat transfer is described entirely by the covariance between volume fraction and fluid temperature fluctuations, also known as the `drift temperature.' Since this quantity is sensitive to variations in \Peclet\; number and mean particle volume fraction, but is minimally sensitive to the ratio of heat capacities, we then leverage scaling arguments and Gene Expression Programming to propose a closure. The resultant model captures the complex dependency of the drift temperature on Pe and $\langle \varepsilon_p \rangle$ and reduces the error in predicting thermal entrance length by 90\% as compared to predictions that neglect heterogeneity. We also note that the proposed model is appropriate for use in both the Reynolds averaged and phase averaged formulations of the heat equations, making it a suitable for use in Euler--Euler codes in which the thermodynamics and hydrodynamics evolve simultaneously.  

\section{Acknowledgements} 
This material is based upon work supported by the National Science Foundation (NSF CAREER, CBET-1846054 and NSF CBET-1904742). The computing resources and assistance provided by the staff of Advanced Research Computing at the University of Michigan, Ann Arbor is greatly appreciated. Additionally, this work used the Extreme Science and Engineering Discovery Environment (XSEDE), which is supported by National Science Foundation grant number ACI-1548562 \citep{XSEDE}. 

\appendix 

\section{Non-dimensionalization of the heat equation}
\label{app:nonndim}
The volume-filtered fluid temperature equation is given by
\begin{equation}
\frac{\partial }{\partial t} \left( \varepsilon_f \rho_f C_{p,f} T_f \right) + \nabla \cdot \left( \varepsilon_f \rho_f C_{p,f} \bm{u}_f T_f \right) = \varepsilon_f \nabla \cdot \left( \kappa_f \nabla T_f \right) + \mathcal{Q}_{\text{inter}}, 
\end{equation}
where, 
\begin{equation}
\mathcal{Q}_{\text{inter}} = - \sum_{i=1}^{N_p} \mathcal{G} \left( \vert \bm{x} - \bm{x}_p^{(i)} \vert \right) q_{\text{heat}}^{(i)},
\end{equation}
and
\begin{equation}
 q_{\text{heat}}^{(i)} = V_p \left \lbrack \frac{6 \kappa_f {\rm Nu}}{d_p^2} \left(T_f \lbrack x_p ^{(i)} \rbrack - T_p^{(i)} \right)\right \rbrack. 
\end{equation}
This results in the following expression, 
\begin{align}
\frac{\partial }{\partial t} \left( \varepsilon_f \rho_f C_{p,f} T_f \right) + &\nabla \cdot \left( \varepsilon_f \rho_f C_{p,f} \bm{u}_f T_f \right) = \nonumber \\
 &\varepsilon_f \nabla \cdot \left( \kappa_f \nabla T_f \right) + \frac{6 \varepsilon_p \kappa_f {\rm Nu}}{d_p^2} \left( T_f - T_p \right).
\end{align}
We first non-dimensionalize temperature using $\theta = (T - T_{p,0})/(T_{f,0}-T_{p,0})$. This yields, 
 \begin{align}
\frac{\partial }{\partial t} \left( \varepsilon_f \rho_f C_{p,f} \theta_f \right) + &\nabla \cdot \left( \varepsilon_f \rho_f C_{p,f} \bm{u}_f \theta_f \right) = \nonumber \\
 &\varepsilon_f \nabla \cdot \left( \kappa_f \nabla \theta_f \right) + \frac{6 \varepsilon_p \kappa_f {\rm Nu}}{d_p^2} \left( \theta_f - \theta_p \right).  
\end{align}
Now, we divide through by $C_{p,f} \rho_f$ and make the change of variable $\hat{\bm{x}} = \bm{x}/d_p$, which gives 
 \begin{align}
\frac{\partial }{\partial t} \left( \varepsilon_f  \theta_f \right) + &\left(\frac{1}{d_p}\right)\frac{\partial}{\partial \hat{\bm{x}}} \left( \varepsilon_f \bm{u}_f \theta_f \right) = \nonumber \\
 &\frac{ \varepsilon_f }{\rho_f C_{p,f} }\left(\frac{1}{d_p}\right)\frac{\partial}{\partial \hat{\bm{x}}} \left( \kappa_f \left(\frac{1}{d_p}\right)\frac{\partial}{\partial \hat{\bm{x}}} \theta_f \right) + \frac{6 \varepsilon_p \kappa_f {\rm Nu}}{\rho_f C_{p,f} d_p^2} \left( \theta_f - \theta_p \right).  
\end{align}
Multiplying by $(d_p/u_{\rm bulk})$ gives rise to 
 \begin{align}
\frac{d_p}{u_{\rm bulk}}\frac{\partial }{\partial t} \left( \varepsilon_f  \theta_f \right) + &\frac{d_p}{u_{\rm bulk}}\left(\frac{1}{d_p}\right)\frac{\partial}{\partial \hat{\bm{x}}} \left( \varepsilon_f \bm{u}_f \theta_f \right) = \nonumber \\
 &\frac{d_p}{u_{\rm bulk}} \varepsilon_f \left(\frac{1}{d_p}\right)\frac{\partial}{\partial \hat{\bm{x}}} \left( \alpha_f \left(\frac{1}{d_p}\right)\frac{\partial}{\partial \hat{\bm{x}}} \theta_f \right) + \frac{6 \varepsilon_p \alpha_f {\rm Nu}}{d_p u_{\rm bulk}} \left( \theta_f - \theta_p \right),  
\end{align}
where we notice that a \Peclet\;number arises in both terms on the right hand side as, 
 \begin{align}
\frac{d_p}{u_{\rm bulk}}\frac{\partial }{\partial t} \left( \varepsilon_f  \theta_f \right) + &\frac{\partial}{\partial \hat{\bm{x}}} \left( \frac{\varepsilon_f \bm{u}_f \theta_f}{u_{\rm bulk}} \right) = \nonumber \\
 & \varepsilon_f\frac{\partial}{\partial \hat{\bm{x}}} \left( \frac{1}{\rm Pe} \frac{\partial}{\partial \hat{\bm{x}}} \theta_f \right) + \frac{6  {\rm Nu}}{\rm Pe}\varepsilon_p \left( \theta_f - \theta_p \right).  
\end{align}
Finally, we define a timescale using $d_p/u_{bulk}$ and a non-dimensional time as $\hat{t} = t/(d_p/u_{bulk})$. Making this change of variable yields 
 \begin{align}
\frac{\partial }{\partial \hat{t}} \left( \varepsilon_f  \theta_f \right) + &\frac{\partial}{\partial \hat{\bm{x}}} \left( \frac{\varepsilon_f \bm{u}_f \theta_f}{u_{bulk}} \right) = \nonumber \\
 & \varepsilon_f\frac{\partial}{\partial \hat{\bm{x}}} \left( \frac{1}{Pe} \frac{\partial}{\partial \hat{\bm{x}}} \theta_f \right) + \frac{6  {\rm Nu}}{\rm Pe}\varepsilon_p \left( \theta_f - \theta_p \right).
\end{align}
Finally, a non-dimensional velocity is defined as $\hat{\bm{u}}_f = \bm{u}_f/u_{bulk}$ and the final non-dimensional equation is given by 
 \begin{align}
\frac{\partial }{\partial \hat{t}} \left( \varepsilon_f  \theta_f \right) + &\frac{\partial}{\partial \hat{\bm{x}}} \left( \varepsilon_f \hat{\bm{u}}_f \theta_f \right) = \nonumber \\
 & \varepsilon_f\frac{\partial}{\partial \hat{\bm{x}}} \left( \frac{1}{Pe} \frac{\partial}{\partial \hat{\bm{x}}} \theta_f \right) + \frac{6  {\rm Nu}}{\rm Pe}\varepsilon_p \left( \theta_f - \theta_p \right),
\end{align}
and can be reorganized as 
 \begin{equation}
{\rm Pe} \frac{\partial }{\partial \hat{t}} \left( \varepsilon_f  \theta_f \right) + {\rm Pe} \frac{\partial}{\partial \hat{\bm{x}}} \left( \varepsilon_f \hat{\bm{u}}_f \theta_f \right) = \varepsilon_f\frac{\partial^2\theta_f }{\partial \hat{\bm{x}}^2}    + 6  {\rm Nu} \varepsilon_p \left( \theta_f - \theta_p \right).
\end{equation}
The particle phase heat equation (Eq.~\eqref{eq:partheat}) can be similarly nondimensionalized. First, the particle phase heat equation is rewritten in the Eulerian sense by conducting a change of frame from the Lagrangian particle heat equation and projecting it to the Eulerian grid. This Eulerian representation is given as,  
\begin{equation}
 \rho_p C_{p,p} \left( \frac{\partial \left(\varepsilon_p  T_p \right)}{\partial t} +   \nabla \cdot \left( \varepsilon_p \bm{u}_p T_p \right) \right)=   \varepsilon_p\kappa_f \nabla^2 T_f + \frac{6 \varepsilon_p \kappa_f \text{Nu}}{d_p^2} \left( T_f - T_p \right). 
\end{equation} 
In the same manner as the fluid heat equation, the particle heat equation is nondimensionalized yielding 
\begin{equation}
     \chi \; {\rm Pe} \frac{\rho_p}{\rho_f}\left \lbrack \frac{{\partial} \left(\varepsilon_p \theta_p \right)}{{\partial} \hat{t}} + \frac{\partial \left( \varepsilon_p \hat{\bm{u}}_p  \theta_p \right)}{\partial \hat{\bm{x}}} \right \rbrack = \varepsilon_p \frac{{\partial}^2 \theta_f}{{\partial} \hat{\bm{x}}^2} + 6 {\rm Nu} \varepsilon_p \left (\theta_f - \theta_p \right) 
\end{equation}
where, $\chi$ is the ratio of heat capacities. 

\section{Development of the 1D heat equation}
\label{app:phaseaverage}

The fluid-phase energy equation is given as 
\begin{equation}
\frac{\partial }{\partial t} \left( \varepsilon_f \rho_f C_{p,f} T_f \right) + \nabla \cdot \left( \varepsilon_f \rho_f C_{p,f} \bm{u}_f T_f \right) = \varepsilon_f \nabla \cdot \left( \kappa_f \nabla T_f \right) + \mathcal{Q}_{\text{inter}}.   
\end{equation}
We conduct Reynolds averaging with respect to the cross-stream directions and time, and assume negligible effects from thermal diffusion. This results in a much simpler expression, given by 
\begin{equation} 
C_{p,f} \rho_f \frac{\deriv}{\deriv x} \langle \varepsilon_f \bm{u}_f T_f \rangle = \langle  \mathcal{Q}_{\text{inter}}\rangle.
\end{equation} 


Note that the interphase heat exchange is given as 
\begin{equation}
\mathcal{Q}_{\text{inter}} = - \sum_{i=1}^{N_p} \mathcal{G} \left( \vert \bm{x} - \bm{x}_p^{(i)} \vert \right) q_{\text{heat}}^{(i)},
\end{equation}
where, 
\begin{equation}
 q_{\text{heat}}^{(i)} = V_p \left \lbrack \frac{6 \kappa_f {\rm Nu}}{d_p^2} \left(T_f \lbrack x_p ^{(i)} \rbrack - T_p^{(i)} \right)\right \rbrack.
\end{equation}
Substituting these definitions into the heat equations and noting that a phase average arises due to the volume fraction in the Reynolds average on the left hand side yields
\begin{align} 
C_{p,f} \rho_f \langle \varepsilon_f \rangle \frac{\deriv}{\deriv x}  \langle {\bm{u}_f {T_f}} \rangle_f  &= -\left \langle   \frac{6 \varepsilon_p \kappa_f {\rm Nu}}{d_p^2} \left(T_f - T_p \right) \right \rangle \\
C_{p,f} \rho_f \langle \varepsilon_f \rangle  \frac{\deriv }{\deriv x} \langle {\bm{u}_f {T_f}} \rangle_f &= - \frac{6 \kappa_f}{d_p^2}\langle \varepsilon_p {\rm Nu} \left(T_f - T_p \right) \rangle 
\end{align} 

Both sides can be simplified further. Working first with the left hand side, we expand $\bm{u}_f$ and $T_f$ using the Phase averaged decompositions, as:
\begin{align}
C_{p,f} \rho_f \langle \varepsilon_f \rangle \frac{\deriv}{\deriv x} \langle {(\langle{\bm{u}_f} \rangle_f+\bm{u}_f^{\prime \prime \prime})( \langle{T_f}\rangle_f + T_f^{\prime \prime \prime})}\rangle_f &=- \frac{6 \kappa_f}{d_p^2}\langle \varepsilon_p {\rm Nu} \left(T_f - T_p \right) \rangle   \\
C_{p,f} \rho_f \langle \varepsilon_f \rangle \frac{\deriv}{\deriv x} \left(\langle{\langle{\bm{u}_f}\rangle_f \langle{T_f}}\rangle_f \rangle_f +\langle{\bm{u}_f^{\prime\prime \prime} T_f^{\prime \prime \prime}}\rangle_f  \right) & =- \frac{6 \kappa_f}{d_p^2}\langle \varepsilon_p {\rm Nu} \left(T_f - T_p \right) \rangle  \\
C_{p,f} \rho_f \langle \varepsilon_f \rangle \frac{\deriv}{\deriv x} \left(\langle{\bm{u}_f}\rangle_f \langle{T_f}\rangle_f +\langle{\bm{u}_f^{\prime \prime \prime} T_f^{\prime \prime \prime}}\rangle_f  \right) &=- \frac{6 \kappa_f}{d_p^2}\langle \varepsilon_p {\rm Nu} \left(T_f - T_p \right) \rangle  \\
C_{p,f} \rho_f \langle \varepsilon_f \rangle \left(\langle{\bm{u}_f}\rangle_f \frac{\deriv \langle{T_f}\rangle_f}{\deriv x} + \frac{\deriv}{\deriv x}\langle{\bm{u}_f^{\prime\prime\prime} T_f^{\prime \prime \prime}}\rangle_f  \right) &= - \frac{6 \kappa_f}{d_p^2}\langle \varepsilon_p {\rm Nu} \left(T_f - T_p \right) \rangle 
\end{align} 

Turning now to the right hand side, we notice that a phase-average with respect to the particle phase is present. 

\begin{align}
    C_{p,f} \rho_f \langle \varepsilon_f \rangle &\left( \langle{\bm{u}_f}\rangle_f \frac{\deriv \langle{T_f}\rangle_f}{\deriv x} + \frac{\deriv}{\deriv x} \langle{\bm{u}^{\prime \prime} T_f^{\prime \prime}}\rangle_f \right) \nonumber \\
    &= -\frac{6 \kappa_f}{d_p^2} \left \langle \varepsilon_p {\rm Nu} (T_f - T_p) \right \rangle \nonumber \\
&= -\frac{6 \kappa_f \langle \varepsilon_p \rangle }{d_p^2} \left \langle {\rm Nu} (T_f - T_p) \right \rangle_p \nonumber \\
&= -\frac{6 \kappa_f \langle \varepsilon_p \rangle }{d_p^2} \left \langle (\langle {\rm Nu} \rangle_p + Nu^{\prime \prime})(\langle T_f \rangle_p + T_f^{\prime \prime} - \langle T_p \rangle_p - T_p^{\prime \prime}) \right \rangle_p \nonumber \\
&= -\frac{6 \kappa_f \langle \varepsilon_p \rangle }{d_p^2} \left \lbrack \langle {\rm Nu} \rangle_p \left(\langle T_f \rangle_p - \langle T_p \rangle_p \right) +  \left( \langle {\rm Nu}^{\prime \prime} T_f^{\prime \prime} \rangle  -  \langle {\rm Nu}^{\prime \prime} T_p^{\prime \prime}\right) \right \rbrack
\end{align}

\section{Reynolds-averaged contributions to phase-averaged terms}
\label{sec:RAvsPA} 
\begin{figure}[ht]
    \centering
    \includegraphics[width = \textwidth]{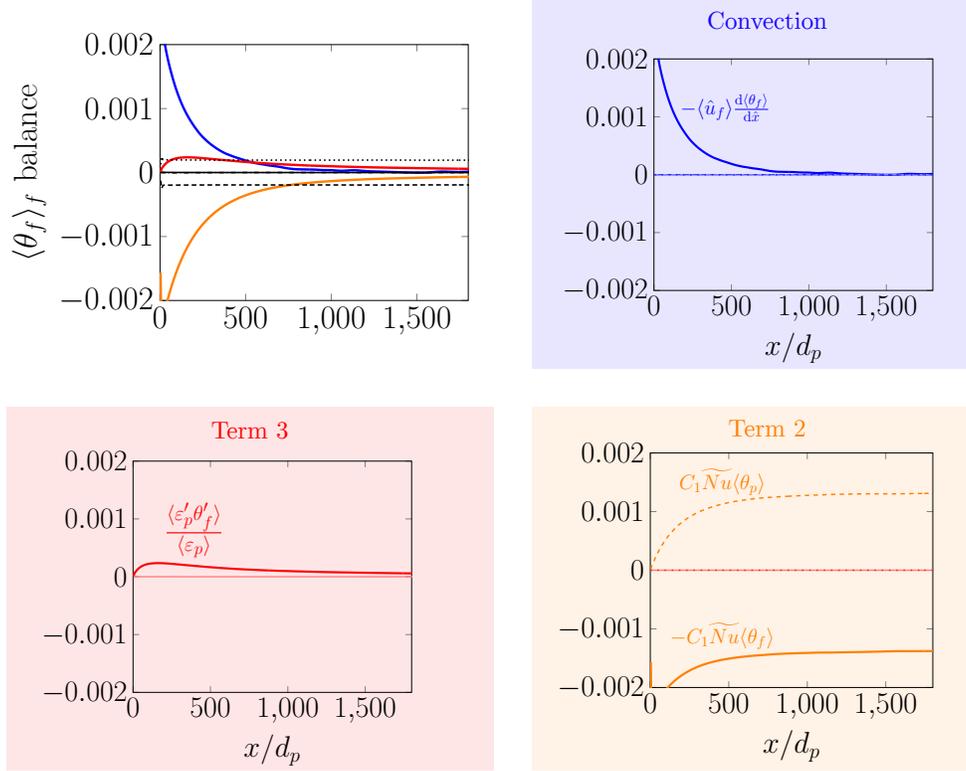}
    \caption{Upper left: Of the terms appearing in the phase-averaged, fluid-phase energy equation, three dominate the thermal behavior: Convection (blue), Term 2 (orange) and Term 3 (red). Terms 4 (black, dotted) and 5 (black, dashed) are nonzero, but balance each other exactly. Colored breakout panels of the three dominate terms detail the contributions to each of these in terms of \emph{Reynolds averaged} quantities. Of the nonzero terms, only $\langle \varepsilon_p^{\prime} \theta_f^{\prime} \rangle$ requires modeling.}
    \label{fig:balance3}
\end{figure}

While the phase-averaged equations have mathematical utility as this significantly reduces the number of terms as compared with Reynolds averaging, proposing models in for the phase-averaged equations in the context of the present study requires additional closure for boundary conditions, since the correlation between temperature and volume fraction fluctuations at the inlet to the thermal domain cannot be known \emph{a priori}. To maintain consistency in boundary conditions in comparing models for correlated and uncorrelated flows, we shift to the Reynolds averaged descriptions of the surviving terms in the phase averaged equations. For the fluid-phase, this exercise results in the expression,
\begin{align}
    &\underbrace{\langle \hat{u}_f \rangle \frac{\deriv \langle \theta_f \rangle}{\deriv \hat{x}} + 
    \frac{\langle \varepsilon_f^{\prime} \hat{u}_f^{\prime} \rangle}{\langle \varepsilon_f \rangle} \frac{\deriv \langle \theta_f\rangle}{\deriv \hat{x}} +
    \frac{\langle \hat{u}_f \rangle}{\langle \varepsilon_f \rangle} \frac{\deriv \langle \varepsilon_f^{\prime} \theta_f^{\prime} \rangle}{\deriv \hat{x}} +
    \frac{\langle \varepsilon_f^{\prime} \hat{u}_f^{\prime} \rangle}{\langle \varepsilon_f \rangle^2}  \frac{\deriv \langle \varepsilon_f^{\prime} \theta_f^{\prime} \rangle}{\deriv \hat{x}}
    }_{\text{Convection}} =  \\ \nonumber
    &- \frac{6 \langle \varepsilon_p \rangle \widetilde{Nu}}{\text{Pe}\langle \varepsilon_f \rangle} \left \lbrack \underbrace{\langle \theta_f \rangle - \langle \theta_p \rangle + \frac{\langle \varepsilon_f^{\prime} \theta_f^{\prime} \rangle}{\langle \varepsilon_f \rangle} -  \frac{\langle \varepsilon_p^{\prime} \theta_p^{\prime} \rangle}{\langle \varepsilon_p \rangle}
    }_{\text{Term 2}} +
    \underbrace{\frac{\langle \varepsilon_p^{\prime} \theta_f^{\prime}\rangle}{\langle \varepsilon_p \rangle} - \frac{\langle \varepsilon_f^{\prime} \theta_f^{\prime}\rangle}{\langle \varepsilon_f\rangle} }_{\text{Term3}}
    \right \rbrack
\end{align} 
where $\widetilde{Nu}$ denotes the Nusslet number computed using the correlation proposed by \citet{Sun2019} and mean quantities as arguments. As shown in the detailed panels of Fig.~\ref{fig:balance3}, all of the unclosed Reynolds averaged terms are null, except for the cross correlation between particle volume fraction and the fluid-phase temperature fluctuations arising from Term 3, as would be expected from \citet{Guo2019}. Thus, the simplified Reynolds averaged equation is given as 
\begin{equation}
  \langle \hat{u}_f \rangle \frac{\deriv \langle \theta_f \rangle}{\deriv \hat{x}}= - \frac{6 \langle \varepsilon_p \rangle \widetilde{Nu}}{\text{Pe}\langle \varepsilon_f \rangle} \left \lbrack \langle \theta_f \rangle - \langle \theta_p \rangle +\frac{\langle \varepsilon_p^{\prime} \theta_f^{\prime}\rangle}{\langle \varepsilon_p \rangle}
    \right \rbrack.
\end{equation} 

This result points to the fact that cross-correlations between volume fraction and temperature shift the phase averaged temperature from the Reynolds averaged temperature (e.g., $\langle \theta_f (\hat{x}) \rangle_f = \langle \theta_f(\hat{x}) \rangle + \langle \varepsilon_f^{\prime} \theta_f^{\prime} \rangle $), where in these configurations the cross correlations are constant with respect to $\hat{x}$. Thus, the model proposed herein is suitable for use in simulations for which the solution variables are phase-averaged \emph{or} Reynolds averaged (see Fig.~\ref{fig:RAvsPA}). This also implies that the proposed model is appropriate for use in a general two-fluid solver in which the hydrodynamics and thermodynamics evolve simultaneously. Of course, in this situation, additional closures are required for the fluid and particle momentum equations in order to capture cross correlations. 

\bibliographystyle{elsarticle-harv} 
\bibliography{CITHT}

\begin{thebibliography}{44}
\expandafter\ifx\csname natexlab\endcsname\relax\def\natexlab#1{#1}\fi
\expandafter\ifx\csname url\endcsname\relax
  \def\url#1{\texttt{#1}}\fi
\expandafter\ifx\csname urlprefix\endcsname\relax\def\urlprefix{URL }\fi

\bibitem[{Abbrecht and Churchill(1960)}]{Abbrecht1960}
Abbrecht, P.~H., Churchill, S.~W., 1960. The thermal entrance region in fully
  developed turbulent flow. {AIC}h{E} Journal 6, 268--273.

\bibitem[{ad~B.~G.~Clarke(2010)}]{Hamdhan2010}
ad~B.~G.~Clarke, I. N.~H., 2010. Determination of thermal conductivity of
  coarse and fine sand soils. Proceedings World Geothermal Congress 2010 Bali,
  Indonesia, 25-29 April 2010.

\bibitem[{Agrawal et~al.(2013)Agrawal, Holloway, Milioli, Milioli, and
  Sundaresan}]{agrawal2013filtered}
Agrawal, K., Holloway, W., Milioli, C.~C., Milioli, F.~E., Sundaresan, S.,
  2013. Filtered models for scalar transport in gas--particle flows. Chemical
  Engineering Science 95, 291--300.

\bibitem[{Beetham and Capecelatro(2019)}]{Beetham2019}
Beetham, S., Capecelatro, J., 2019. Biomass pyrolysis in fully-developed
  turbulent riser flow. Renewable Energy 140, 751--760.

\bibitem[{Beetham et~al.(2021)Beetham, Fox, and Capecelatro}]{Beetham2021}
Beetham, S., Fox, R.~O., Capecelatro, J., 2021. Sparse identification of
  multiphase turbulence closures for coupled fluid-particle flows. Journal of
  Fluid Mechanics 914, A11.

\bibitem[{Capecelatro and Desjardins(2013)}]{Capecelatro2013}
Capecelatro, J., Desjardins, O., 2013. An {E}uler--{L}agrange strategy for
  simulating particle-laden flows. Journal of Computational Physics 238, 1--31.

\bibitem[{Capecelatro et~al.(2014)Capecelatro, Desjardins, and
  Fox}]{capecelatro2014cit}
Capecelatro, J., Desjardins, O., Fox, R.~O., 2014. Numerical study of
  collisional particle dynamics in cluster-induced turbulence. Journal of Fluid
  Mechanics 747, R2 1--13.

\bibitem[{Capecelatro et~al.(2015)Capecelatro, Desjardins, and
  Fox}]{capecelatro2015}
Capecelatro, J., Desjardins, O., Fox, R.~O., 2015. On fluid-particle dynamics
  in fully-developed cluster-induced turbulence. Journal of Fluid Mechanics
  780, 578--635.

\bibitem[{Cundall and Strack(1979)}]{Cundall1979}
Cundall, P.~A., Strack, O. D.~L., 1979. A discrete numerical model for granular
  assemblies. Geotechnique 29~(1), 47--65.

\bibitem[{Desjardins et~al.(2008)Desjardins, Blanquart, Balarac, and
  Pitsch}]{desjardins2008high}
Desjardins, O., Blanquart, G., Balarac, G., Pitsch, H., 2008. High order
  conservative finite difference scheme for variable density low {M}ach number
  turbulent flows. Journal of Computational Physics 227~(15), 7125--7159.

\bibitem[{Ebert and Glicksman(1993)}]{Ebert1993}
Ebert, T.~A., Glicksman, L.~R., 1993. Determination of particle and gas
  convective heat trasnfer components in a circulating fluidized bed. Chemical
  Engineering Science 48, 2179--2188.

\bibitem[{Fox(2014)}]{fox2014}
Fox, R.~O., 2014. On multiphase turbulence models for collisional
  fluid--particle flows. Journal of Fluid Mechanics 742, 368--424.

\bibitem[{Goyal et~al.(2018)Goyal, Desjardins, Pepiot, and
  Capecelatro}]{capecelatro2018biomass}
Goyal, H., Desjardins, O., Pepiot, P., Capecelatro, J., 2018. A computational
  study of the effects of multiphase dynamics in catalytic upgrading of biomass
  pyrolysis vapor. {AIC}h{E} Journal 64, 3341--3353.

\bibitem[{Guo and Capecelatro(2019)}]{Guo2019}
Guo, L., Capecelatro, J., 2019. The role of clusters on heat transfer in
  sedimenting gas-solid flows. International Journal of Heat and Mass Transfer
  132, 1217--1230.

\bibitem[{Hasan(2012)}]{Hasan2012}
Hasan, B.~O., 2012. Heat transfer analysis in thermal entrance region under
  turbulent flow conditions. Asia-{P}acific Journal of Chemical Engineering 8,
  578--592.

\bibitem[{Incropera(2011)}]{incropera_2011}
Incropera, F.~P., 2011. Fundamentals of heat and mass transfer. Wiley.

\bibitem[{Issangya et~al.(2000)Issangya, Grace, Bai, and Zhu}]{Issangya2000}
Issangya, A.~S., Grace, J., Bai, D., Zhu, J., 2000. Further measurements of
  flow dynamics in a high-density circulating fluidized bed riser. Powder
  Technology 111, 104--113.

\bibitem[{Jofre et~al.(2020)Jofre, del Rosario, and Iaccarino}]{Jofre2020}
Jofre, L., del Rosario, Z., Iaccarino, G., 2020. Data-driven dimensional
  analysis of heat transfer in irradiated particle-laden turbulent flow.
  International Journal of Multiphase Flow 125, 1--15.

\bibitem[{Lathouwers and Bellan(2001)}]{lathouwers2001modeling}
Lathouwers, D., Bellan, J., 2001. Modeling of dense gas--solid reactive
  mixtures applied to biomass pyrolysis in a fluidized bed. Int. J. Multiphase
  Flow 27~(12), 2155--2187.

\bibitem[{Lee(1968)}]{Lee1968}
Lee, Y., 1968. Turbulent heat transfer from the core tube in thermal entrance
  region of concentric annuli. International Journal of Heat and Mass Transfer
  11, 509--522.

\bibitem[{Lei et~al.(2020)Lei, Zhu, and Luo}]{Lei2020}
Lei, H., Zhu, L.-T., Luo, Z.-H., 2020. Study of filtere interphase heat
  transfer using highly resolved {CFD}-{DEM} simulations. A{IC}h{E} Journal:
  Particle Techology and Fluidization, 1--12.

\bibitem[{Louge et~al.(1993)Louge, Yusof, and Jenkins}]{Louge1993}
Louge, M., Yusof, J.~M., Jenkins, J.~T., 1993. Heat transfer in the pneumatic
  transport of massive particles. International Journal of Heat and Mass
  Transfer 36, 265--275.

\bibitem[{Lube et~al.(2020)Lube, Breard, Esposti-Ongaro, Dufek, and
  Brand}]{Lube2020}
Lube, G., Breard, E. C.~P., Esposti-Ongaro, T., Dufek, J., Brand, B., 2020.
  Multiphase flow behaviour and hazard prediction ofpyroclastic density
  currents. Nature Reviews Earth and Environment 1, 348–365.

\bibitem[{Mehos et~al.(2017)Mehos, Turchi, Vidal, Wagner, Ma, Ho, Kolb,
  Andraka, and Kruizenga}]{Mehos2017}
Mehos, M., Turchi, C., Vidal, J., Wagner, M., Ma, Z., Ho, C., Kolb, W.,
  Andraka, C., Kruizenga, A., 2017. Concentrating solar power gen3
  demonstration roadmap. Technical report, National Renewable Energy Laboratory
  (NREL), Golden, CO, United States.

\bibitem[{Miller et~al.(2014)Miller, Syamlal, Mebane, Storlie, Bhattacharyya,
  Sahinidis, Agarwal, Tong, Zitney, Sarkar, Sun, Sundaresan, Ryan, Engel, and
  Dale}]{miller2014carbon}
Miller, D.~C., Syamlal, M., Mebane, D.~S., Storlie, C., Bhattacharyya, D.,
  Sahinidis, N.~V., Agarwal, D., Tong, C., Zitney, S.~E., Sarkar, A., Sun, X.,
  Sundaresan, S., Ryan, E., Engel, D., Dale, C., 2014. Carbon capture
  simulation initiative: a case study in multiscale modeling and new
  challenges. Annual Review of Chemical and Biomolecular Engineering 5,
  301--323.

\bibitem[{Morris et~al.(2016)Morris, Ma, Pannala, , and Hrenya}]{Morris2016}
Morris, A.~B., Ma, Z., Pannala, S., , Hrenya, C.~M., 2016. Simulations of heat
  transfer to solid particles flowing through an array of heated tubes. Solar
  Energy 130, 101--115.

\bibitem[{Noymer and Glicksman(1998)}]{Noymer1998}
Noymer, P.~D., Glicksman, L.~R., 1998. Cluster motion and particle-convective
  heat trasnfer at the wall of a circulating fluidized bed. International
  Journal of Heat and Mass Transfer 41, 147--158.

\bibitem[{Peng et~al.(2019)Peng, Kong, Zhou, Sun, Passalacqua, Subramaniam, and
  Fox}]{Peng2019}
Peng, C., Kong, B., Zhou, J., Sun, B., Passalacqua, A., Subramaniam, S., Fox,
  R.~O., 2019. Implementation of pseudo-turbulence closures in an
  {E}ulerian–{E}ulerian two- fluid model for non-isothermal gas–solid flow.
  Chemical Engineering Science 207, 663--671.

\bibitem[{Pielichowska and Pielichowski(2014)}]{Pielichowska2014}
Pielichowska, K., Pielichowski, K., 2014. Phase change materials for thermal
  energy storage. Progreses in Materials Science 65, 67--123.

\bibitem[{Pouransari and Mani(2017{\natexlab{a}})}]{Pouransari2017}
Pouransari, H., Mani, A., 2017{\natexlab{a}}. Effects of preferential
  concentration on heat transfer in particle- based solar receivers. Journal of
  Solar Energy Engineering 139, 021008.

\bibitem[{Pouransari and Mani(2017{\natexlab{b}})}]{pouransari2017effects}
Pouransari, H., Mani, A., 2017{\natexlab{b}}. Effects of preferential
  concentration on heat transfer in particle-based solar receivers. Journal of
  Solar Energy Engineering 139~(2).

\bibitem[{Qiu et~al.(2000)Qiu, Murashov, and White}]{Qiu2000}
Qiu, L., Murashov, V., White, M.~A., 2000. Zeolite {4A}: heat capacity and
  thermodynamics properties. Solid State Sciences 2, 841--846.

\bibitem[{Rauchenzauner and Schneiderbauer(2020)}]{Rauchenzauner2020}
Rauchenzauner, S., Schneiderbauer, S., 2020. A dynamic spatially averaged
  two-fluid model for heat transport in moderately dense gas–particle flows.
  Physics of Fluids 32, 063307:1--20.

\bibitem[{Searson(2009)}]{Searson2009}
Searson, D., 2009. Gptips: Genetic programming \& symbolic regression for
  {MATLAB}. http://gptips.sourceforge.net.

\bibitem[{Shah et~al.(2016)Shah, Utikar, Pareek, Evans, and Joshi}]{Shah2016}
Shah, M.~T., Utikar, R.~P., Pareek, V.~K., Evans, G.~M., Joshi, J.~B., 2016.
  Computational fluid dynamic modelling of {FCC} riser: A review. Chemical
  Engineering Research and Design 111, 403--448.

\bibitem[{Sparrow et~al.(1957)Sparrow, Hallman, and Siegel}]{Sparrow1957}
Sparrow, E.~M., Hallman, T.~M., Siegel, R., 1957. Turbulent heat transfer in
  the thermal entrance region of a pipe with uniform heat flux. Applied
  Scientific Research, Section A 7, 37--52.

\bibitem[{Sun et~al.(2016)Sun, Tenneti, Subramaniam, and Koch}]{Sun2016}
Sun, B., Tenneti, S., Subramaniam, S., Koch, D.~L., 2016. Pseudo-turbulent heat
  flux and average gas–phase conduction during gas–solid heat transfer:
  flow past random fixed particle assemblies. Journal of Fluid Mechanics 798,
  299--349.

\bibitem[{Sun and Zhu(2019)}]{Sun2019}
Sun, Z., Zhu, J., 2019. A consolidated flow regime map of upward gas
  fluidization. AIChE Journal 65, 1--15.

\bibitem[{Tenneti and Subramaniam(2011)}]{Tenneti2011}
Tenneti, S., Subramaniam, S., 2011. Drag law for monodisperse gas-solid systms
  using particle-resolved direct numerical simulation of flow past fixed
  assemblies of spheres. International Journal of Multiphase Flow 37~(9),
  1072--1092.

\bibitem[{Towns et~al.(2014)Towns, Cockerill, Dahan, Foster, Gaither, Grimshaw,
  Hazlewood, Lathrop, Lifka, Peterson, Roskies, Scott, and
  Wilkins-Diehr}]{XSEDE}
Towns, J., Cockerill, T., Dahan, M., Foster, I., Gaither, K., Grimshaw, A.,
  Hazlewood, V., Lathrop, S., Lifka, D., Peterson, G.~D., Roskies, R., Scott,
  J.~R., Wilkins-Diehr, N., 2014. Xsede: Accelerating scientific discovery.
  Computing in Science \& Engineering 16, 62--74.

\bibitem[{Welty et~al.(2019)Welty, Rorrer, and Foster}]{Welty2019}
Welty, J.~R., Rorrer, G.~L., Foster, D.~G., 2019. Fundamentals of momentum,
  heat, and mass transfer. John Wiley \& Sons, Inc.

\bibitem[{Wilson et~al.(1978)Wilson, Sparks, Huang, and Watkins}]{Wilson1978}
Wilson, L., Sparks, R. S.~J., Huang, T.~C., Watkins, N.~D., 1978. The control
  of volcanic column heights by eruption energetics and dynamics. Journal of
  Geophysical Research 83, 1829--1836.

\bibitem[{Xu and Zhang(2002)}]{Xu2002}
Xu, Z., Zhang, Y., 2002. Quench rates in air, water, and liquid nitrogen, and
  interference of temperature in volcanic eruption columns. Earth and Planetary
  Science Letters 200, 315--330.

\bibitem[{Yousefi et~al.(2021)Yousefi, Ardekani, Picano, and
  Brandt}]{Yousefi2021}
Yousefi, A., Ardekani, M., Picano, F., Brandt, L., 2021. Regimes of heat
  transfer in finite-size particle suspensions. International Journal of Heat
  and Mass Transfer 177, 121514.

\end{thebibliography}

\end{document}